\documentclass[final]{siamltex}
\usepackage{euscript,amsmath,amssymb,amsfonts,graphicx,bm,cite,appendix}
  \usepackage{paralist}
  \usepackage{epstopdf}
  \usepackage{graphics} 
 \usepackage[colorlinks=true]{hyperref}
 \hypersetup{urlcolor=blue, citecolor=red}

\newcommand{\Rset}{{\mathbb R}}
\newcommand{\mc}{\mathcal}
\newcommand{\pd}{\partial}
\newcommand{\x}{{\mathbf x}}
\newcommand{\y}{{\mathbf y}}
\newcommand{\rr}{{\mathbf r}}
\newcommand{\ab}{{\mathbf a}}
\newcommand{\e}{{\rm e}}
\newcommand{\bd}{{\boldsymbol \Delta}}
\newcommand{\ve}{\varepsilon}

\newcommand{\argmax}{{\rm argmax}}

\title{Stochastic motion of bumps in planar neural fields}

\author{Daniel Poll\thanks{Department of Mathematics, University of Houston, Houston, Texas 77204 ({\tt dbpoll@math.uh.edu})}\and Zachary P. Kilpatrick\thanks{Department of Mathematics, University of Houston, Houston, Texas 77204, USA ({\tt zpkilpat@math.uh.edu})}}

\date{\today}

\begin{document}

\maketitle
\newcommand{\slugmaster}{%
\slugger{MMedia}{xxxx}{xx}{x}}

\begin{abstract}
We analyze the effects of spatiotemporal noise on stationary pulse solutions (bumps) in neural field equations on planar domains. Neural fields are integrodifferential equations whose integral kernel describes the strength and polarity of synaptic interactions between neurons at different spatial locations of the network. Fluctuations in neural activity are incorporated by modeling the system as a Langevin equation evolving on a planar domain. Noise causes bumps to wander about the domain in a purely diffusive way. Utilizing a small noise expansion along with a solvability condition, we can derive an effective stochastic equation describing the bump dynamics as two-dimensional Brownian motion. The diffusion coefficient can then be computed explicitly. We also show that weak external inputs can pin the bump so it no longer wanders diffusively. Inputs reshape the effective potential that guides the dynamics of the bump position, so it tends to lie near attractors which can be single points or contours in the plane. Perturbative analysis shows the bump position evolves as a multivariate Ornstein-Uhlenbeck process whose relaxation constants are determined by the shape of the input. Our analytical approximations all compare well to statistics of bump motion in numerical simulations.

\end{abstract}

\section{Introduction} Persistent and localized neural activity has been observed in a number of experiments probing mechanisms of sensation \cite{weber04,katzner09} and memory \cite{moser08,goldmanrakic95}. In particular, recordings from prefrontal cortex during spatial working memory tasks link the spatiotemporal dynamics of neural activity to an animal's resulting behavior \cite{funahashi89}. When a monkey is trained to recall the specific position of a cue, the network location of persistent activity encodes the corresponding cue location. Any displacement of the persistent activity from its initial location is reflected by errors the monkey makes in recalling the cue's position \cite{wimmer14}. In a related way, persistent activity in the entorhinal cortex \cite{hafting05} and hippocampus \cite{burgess02} can store an animal's physical location in its environment, constituting an internal ``global positioning system" \cite{abbott14}. While most studies of spatial working memory tend to focus on recalling an analog variable in one-dimension \cite{wang01}, networks performing such spatial navigation have been shown to represent space in two \cite{mcnaughton06} and even three dimensions \cite{yartsev13}.

Continuum neural field equations on planar domains are a well accepted model of spatially structured neuronal activity evolving on the surface of in vitro and in vivo brain tissue \cite{wang10,bressloff12}. However, analyses of such equations tend to focus exclusively on deterministic spatiotemporal dynamics \cite{laing02,folias05,kilpatrick10,owen07}, ignoring the impact of the nervous system's fluctuations \cite{faisal08}. Extending his seminal work on the stability of stationary bump solutions in one-dimension \cite{amari77}, Amari was the first to explore the dynamics of spatially coherent activity in planar neural fields, focusing on $D_n$-symmetric azimuthal instabilities of radially symmetric bumps \cite{amari78,amari14}. Several authors have extended this work, studying the solutions that emerge from such symmetry breaking instabilities such as multi-bumps \cite{owen07}, breathers \cite{folias05}, and labyrinthine patterns \cite{coombes12}. In addition, $D_1$-symmetric modes of radially symmetric bumps can be destabilized by incorporating an additional negative feedback variable into the evolution equations, modeling spike frequency adaptation \cite{coombes07} or synaptic depression \cite{bressloff11}. In this case, bumps no longer remain stationary when deformed by shifting perturbations. Indeed, these previous studies have shed light on how the architecture and parameters of neural field models shape the dynamics of their solutions.

Noise fluctuations have a considerable impact on the large scale dynamics of neuronal activity \cite{fox07}. A wide variety of previous modeling studies have explored how noise and chaos affect bulk neuronal activity in high-dimensional spiking networks \cite{vreeswijk96,destexhe06,benayoun10}. However, recent work has also shown that incorporating weak noise into neural field models can qualitatively reshape their long term dynamics \cite{bressloff12}. Studies of this sort proceed by casting the model in terms of a Langevin equation evolving on a spatial domain subject to spatiotemporal noise. Weak additive noise can shift the threshold of Turing bifurcations, beyond which stationary spatial patterns form \cite{hutt08}. Furthermore, noise leads to diffusive motion of fronts \cite{bressloff12b} and bumps \cite{kilpatrick13}, so they execute a random walk about their mean position. Analysis and simulation of these phenomena is inspired by studies of stochastic front propagation in reaction-diffusion PDE models \cite{armero98,panja04}. Complementary to these perturbative approaches, Kruger and Stannat have recently developed a multiscale decomposition of solutions to stochastic neural field models which allows for a rigorous treatment of the existence and stability of solutions \cite{kruger14}. Furthermore, Faugeras and Ingles have addressed the issue of well-posedness of solutions to stochastic neural fields \cite{faugeras14}. Neural fields and reaction-diffusion models tend to possess translation symmetry, and, as a result, their spatially structured solutions are neutrally stable to spatial translations that preserve their profile. However, if the translation symmetry of a neural field equation is broken by spatial heterogeneity, its solutions will be linearly stable to translating perturbations. In this case, they will tend to stay close to their initial position in the presence of weak noise perturbations \cite{bressloff12b,kilpatrick13}.

We extend these previous findings by exploring the stochastic motion of bumps in planar neural fields, reaching beyond current work that tends to explore one-dimensional domains \cite{hutt08,bressloff12}. Our principle finding is that weak noise causes stationary bumps to execute two-dimensional Brownian motion, which can be quantified with an effective diffusion coefficient. Our derivations require the enforcement of a solvability condition for the linearization of a nonlinear Langevin equation on the plane. Since bump position represents a memory of an initial condition, we are interested in what features of the model shape the long term diffusion of the bump. Thus, we also analyze the impact of external inputs on stochastic bump motion, finding they tend to pin the bumps to their peaks. The paper proceeds by first introducing the stochastic neural field model on the plane in section \ref{smodel}. We then begin section \ref{wbr2} by reviewing previous analyses of existence and stability of radially symmetric bumps \cite{owen07}, providing intuition for the impact of noise perturbations on bump position. Subsequently, we derive an effective equation for the dynamics of radially symmetric bumps subject to weak noise, deriving an effective diffusion coefficient for the variance of the bump's position. In section \ref{stimpin}, we extend these results by incorporating the effects of an external input on the stochastic dynamics of bumps. Inputs stabilize bumps to translating perturbations, so the position of noise-driven bumps evolves approximately as an Ornstein-Uhlenbeck process in two-dimensions.

\section{Stochastic neural fields on planar domains}
\label{smodel}

We will analyze the effective diffusion of stationary bump solutions in planar neural field equations with noise. A wide variety of previous studies have analyzed the existence and stability of bumps in two-dimensional neural fields. Studies in bounded domains are often eased by the fact that solutions can be decomposed into finite expansions of basis functions \cite{bressloff03,faugeras09}. There are also several successful analysis of the existence and stability of bumps on unbounded domains, such as $\Rset^2$ \cite{laing03,owen07}. As a starting point, we will review some previous analyses of bump solutions in the planar neural field model \cite{taylor99,werner01,owen07,amari14}
\begin{equation} \label{model}
\frac{ \partial u(\x,t) }{ \partial t} = - u(\x,t) + \int_{\Rset^2} w(\x-\y) f( u(\y,t)) \; d \y,
\end{equation}
where $u ( \x ,t) $ denotes the total synaptic input to the neural field at the position $\x = (x_1,x_2) \in \Rset^2$. The integral term describes the synaptic connectivity of the network, so that $w (\x - \y) $ describes the strength (amplitude) and polarity (sign) of connectivity from neurons at location $\y$ to neurons at location $\x$. In our analysis, we utilize a sum of $N+1$ modified Bessel functions as our typical weight function $w$ to demonstrate the results we derive
\begin{equation}  \label{weight}
w(\x - \mathbf{y}) = \sum_{j=0}^N c_j K_0( \alpha_j ||\mathbf{x-y} || ), 
\end{equation}
where $K_v$ is a modified Bessel function of the second kind of order $v$. The constants $c_j$ and $\alpha_j$ scale the amplitude and spatial decay of the $j$th term in the Bessel function sum. One advantage of the weight function (\ref{weight}) is that integrals arising in (\ref{model}) can be computed analytically with the aid of Hankel transforms \cite{laing03,folias05,owen07}. Note that $||.||$ denotes the standard Euclidean 2-norm
$$
|| \x|| = \sqrt{ x_1^2 + x_2^2}
$$
so that $w$ of the form (\ref{weight}) will be radially symmetric. The function $f$ denotes the firing rate of the model, which is a representation of the fraction of  total active neurons, $0 \leq f \leq 1$. Experimental data suggests $f$ should roughly be a sigmoidal function \cite{wilson72}
\begin{equation} \label{fire}
f(u) = \frac{1}{1 + e^{- \gamma (u - \kappa) } }
\end{equation}
where $\gamma$ is the gain and $\kappa$ is the threshold. The effective equations we derive for the stochastic motion of bumps will hold for general firing rate functions $f$, but we consider the high gain limit $\gamma \rightarrow \infty$ of (\ref{fire}) to compute the resulting formulas explicitly. In this case \cite{amari77,amari14}
\begin{align} \label{Hrate}
f(u) := H(u - \kappa) = \begin{cases}
                       1: &u \geq \kappa, \\
                       0: &u < \kappa,
                      \end{cases}
\end{align}
so the firing rate function is a Heaviside function.

The planar neural field equation (\ref{model}) in the absence of noise has been studied extensively, demonstrating a wide variety of neural patterns \cite{laing03,owen07,coombes12,amari14}. However, our main concern is the impact of noise on stationary bump solutions of (\ref{model}). We focus on a model that incorporates additive noise into a planar neural field, an extension of recent studies that have explored how noise shapes spatiotemporal dynamics of neural fields in one-dimensional domains \cite{laing01,brackley07,hutt08,bressloff12}. The model takes the form of a Langevin equation on the plane $\Rset^2$ forced by a spatiotemporal noise process
\begin{equation} \label{modeln}
du(\x,t) = \bigg( - u(\x,t) + \int_{ \Rset^2} w(\x-\y) f(u(\y,t)) \; d\y \bigg) dt + \ve^{1/2} dW(\x,t).
\end{equation}
The term $dW(\x,t)$ is the increment of a spatially varying Wiener process with spatial correlations defined
\begin{equation} \label{noise}
\langle dW(\x,t) \rangle = 0, \; \; \langle dW(\x,t) dW(\y,s) \rangle = C(\x - \y ) \delta (t-s) dt ds,
\end{equation}
so that $\ve$ describes the intensity of the noise, which we assume to be weak ($\ve \ll 1$). The function $C( \x - \y) $ describes the spatial correlation in each noise increment between two points $\x, \y \in \Rset^2$. The spatial correlation function $C( \x - \y )$ can be related directly to the spatial filter ${\mc F}( \x - \y)$. First, we note that $d W (\x, t)$ can be defined by convolving a spatially white noise process $d {\mc Y}(\x ,t)$, satisfying $\langle d {\mc Y}(\x, t) \rangle = 0 $ and $\langle d {\mc Y} ( \x , t) d {\mc Y}( \y ,s) \rangle = \delta ( \x - \y ) \delta ( t-s) dt ds$, with the filter ${\mc F}(\x - \y)$, so
\begin{align*}
d W( \x , t) = \int_{\Omega} {\mc F}(\x - \y ) d {\mc Y}(\y,t) d \y.
\end{align*}
Thus, we can determine how the variance of $d W ( \x ,t)$ depends on the filter ${\mc F}( \x - \y )$ by computing
\begin{align}
\langle d W (\x ,t) d W( \y, t) \rangle &= \left\langle \int_{\Omega} {\mc F}(\x - \x' ) d {\mc Y}(\x',t) d \x' \int_{\Omega} {\mc F}(\y - \y' ) d {\mc Y}(\y',t) d \y' \right\rangle  \nonumber \\
&=  \int_{\Omega} \int_{\Omega} {\mc F}(\x - \x' ) {\mc F}(\y - \y' ) \langle d {\mc Y}(\x',t) d {\mc Y}(\y',t) \rangle d \y' d \x'   \nonumber  \\
&=  \int_{\Omega} \int_{\Omega} {\mc F}(\x - \x' ) {\mc F}(\y - \y' ) \delta (\x' - \y') d \y' d \x' \delta (t-s) dt ds  \nonumber \\
\langle d W (\x ,t) d W( \y, t) \rangle &= \int_{\Omega} {\mc F}(\x - \x' ) {\mc F}(\y - \x' ) d \x'  \delta (t-s) dt ds \nonumber \\ &= C( \x - \y) \delta (t-s) dt ds,  \label{dWfilt}
\end{align}
so
\begin{align}
C( \x - \y ) = \int_{\Omega} {\mc F}(\x - \x' ) {\mc F}(\y - \x' ) d \x'.  \label{corfilt}
\end{align}
The last equality in (\ref{dWfilt}) holds due to our definition of $d W( \x ,t)$. Note that (\ref{corfilt}) implies that $C(\x - \y)$ should be an even symmetric function, since the arguments of both functions ${\mc F}( \x )$ can be exchanged. In other words, the points $\x$ and $\y$  in $C( \x - \y)$ can also be exchanged.

%\begin{equation}
%\begin{aligned}
%&\int_\Omega \int_\Omega F(\mathbf{x-x'})F(\mathbf{y-y'}) d \mathcal{Y}(\x,t) d \mathcal{Y} (\mathbf{y},s) d \mathbf{x'} d \mathbf{y'}\\
%= &\int_\Omega \int_\Omega F(\mathbf{x-x'})F(\mathbf{y-y'}) \delta(t-s) \delta(\mathbf{x' - y'}) d \mathbf{x'} d \mathbf{y'} dt ds \\
%= &\int_\Omega F(\mathbf{x-x'})F(\mathbf{y-x'}) d \mathbf{x'}  \; \delta(t-s) dt ds \\
%= &\;  C(\mathbf{x-y}) \delta(t-s)
%\end{aligned}
%\end{equation}
%where $F$ is the respective noise filter and $\delta$ is the usual dirac delta function. This definition induces an evenness property, as our filter $F$ commutes with itself, which we would expect for two points to be considered correlated.  
%\\ \\
As an example, consider the noise filter ${\mc F}(\x) = \cos(x_1) + \cos(x_2) + \sin(x_1) + \sin(x_2)$. To compute the associated spatial correlation function, we must restrict integration to a compact domain $\Omega \subset \Rset^2$ of size $2 M \pi \times 2 M \pi$ ($M \in \mathbb{Z}$). Note, in numerical simulations, we are forced to do so anyway. By utilizing (\ref{corfilt}), we find
\begin{align}
C( \x - \y ) =& \int_{\Omega} {\mc F}(\x - \x'){\mc F}(\y-\x') d \x' \nonumber \\
= & \int_{\Omega} \sum_{j=1}^2  \left[  \cos(x_j-x_j') + \sin(x_j-x_j') \right] \sum_{j=1}^2 \left[  \cos(y_j-x_j') +  \sin(y_j-x_j') \right] d \x \nonumber\\ 
= & \int_{\Omega} \sum_{j=1}^2 \left[  \cos^2 (x_j') \cos (x_j) \cos (y_j) +  \sin^2 (x_j') \sin (x_j) \sin (y_j) \right] d \x \nonumber \\
= & 2 M^2 \pi^2 \sum_{j=1}^2 \left[  \cos (x_j) \cos (y_j) +  \sin (x_j) \sin (y_j) \right] \nonumber \\
C( \x - \y ) = & 2 M^2 \pi^2 \left[ \cos (x_1 - y_1) + \cos (x_2 - y_2) \right].  \label{derivecc}
\end{align}
Note that the size of the domain is controlled by $M$ and this also scales the relationship between the noise filter and correlation function, which is important to remember when comparing numerical to analytical results. We will employ this canonical spatial noise filter in our explorations of stochastic bump motion.

%\begin{equation}
%\begin{aligned}
%&\int_{\Omega} {\mc F}(\mathbf{x-x'}){\mc F}(\mathbf{y-x'}) d \mathbf{x'} = \sum_{j=1}^2\cos(x_j)\cos(y_j) + \sin(x_j)\sin(y_j) \\
%= & \; \cos(x_1 - y_1) + \cos(x_2 - y_2) = C(\mathbf{x - y} )
%\end{aligned}
%\end{equation}
%which we will use as part of an example in Section 3.3.

\section{Wandering bumps in $\Rset^2$} \label{wbr2}
We begin by studying bumps in the unbounded domain $\Omega = \mathbb{R}^2$, first in the absence of noise (\ref{model}) and then in the presence of additive noise (\ref{modeln}). Recent studies have shown traveling waves and bumps in stochastic neural fields wander diffusively about their mean position, but these analyses have focused on one-dimensional domains\cite{bressloff12,kilpatrick13}. Our analysis will allow us to approximate the diffusion coefficient of a bump driven by noise in a two-dimensional domain. Since we are exploring the model (\ref{modeln}) in $\Rset^2$, we can utilize Hankel transforms to compute integral terms \cite{folias05}.

\subsection{Existence and stability of bumps}
To begin, we review prior results constructing rotationally symmetric stationary bump solutions in the noise-free system (\ref{model}) \cite{taylor99,owen07,amari14}. Specifically, we employ the assumption that the weight function (\ref{weight}) is rotationally symmetric to look for stationary solutions of the form $u(\x,t) : = U(\x) = U(|| \x ||)$. In this case, the neural field equation (\ref{model}) simplifies to 
\begin{equation} \label{stationary}
U(|| \x ||) = \int_{ \mathbb{R}^2} w( \x - \y ) f( U( \y )) d \y.
\end{equation}
By changing variables to polar coordinates $\x = (x_1,x_2) \mapsto \rr = (r, \theta)$, we can convert (\ref{stationary}) to a double integral of the form
\begin{equation}
 U(r) = \int_0^{2 \pi} \int_0^{\infty} w(| \mathbf{r - r'}|) f(U(\rr')) r' dr' d \theta'.  \label{polarU}
\end{equation}
Note that if we assume a Heaviside firing rate function (\ref{Hrate}), then (\ref{polarU}) becomes
\begin{align}
 U(r) = \int_0^{2 \pi} \int_0^{a} w(| \mathbf{r - r'}|) r' dr' d \theta',  \label{polhu}
\end{align}
where $U( \rr ) > \kappa$ when $r<a$ and $U( \rr ) < \kappa$ when $r>a$, so that $r \equiv a$ defines the boundary of the bump. An advantage of utilizing a Heaviside firing rate function (\ref{Hrate}) is that the stability of the bump can be probed by analyzing the dynamics of the boundary. When we analyze the stochastic motion of the bump, we will also derive effective equations by focusing on perturbations of the bump boundary by spatiotemporal noise. We can evaluate the integral in (\ref{polhu}) using Hankel transform and Bessel function identities, as in \cite{folias05,owen07}
\begin{align}
U(r) = \int_0^{\infty} \widehat{w}(\rho) J_0 ( r \rho ) J_1(a \rho) d \rho,  \label{Ubess}
\end{align}
where $J_{\nu}(z)$ is a Bessel function of the first kind of order $\nu$ and the Hankel transform is defined
\begin{align*}
\widehat{w}( \rho ) = \int_{ \Rset^2} \e^{i {\bf h} \cdot \rr } w( \rr ) d \rr,
\end{align*}
where $\rho = ||{\bf h}||$.

\begin{figure}
\begin{center} \includegraphics[width=13cm]{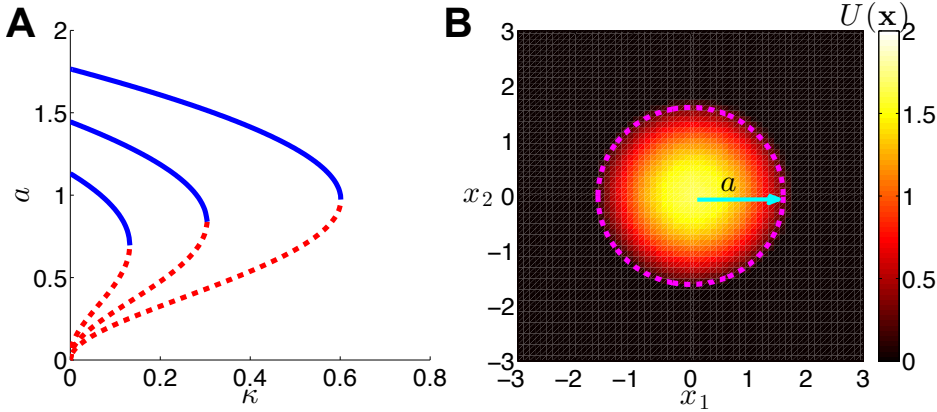} \end{center}
\caption{Stationary bump solutions $u(\x,t) = U(r)$ to the deterministic model (\ref{model}) with weight function (\ref{weight}) and Heaviside firing rate (\ref{Hrate}). ({\bf A}) A wider stable (solid) and narrower unstable (dashed) branch of solutions to $U(a) = \kappa$ emerges from a saddle-node bifurcation at a critical value of $\kappa$ above which no solutions exist. Weight function parameters are (left to right) $[c_1, c_2, c_3, c_4 ] = [10/9,-10/9,-1/3,1/3]; [4/3,-4/3,-2/5,1/3]; [5/3,-5/3,-1/2,1/2]$. ({\bf B}) An example of a stationary bump $U(\x)$ for the parameters $\kappa = 0.2$ and $[c_1, c_2, c_3, c_4 ]=[5/3,-5/3,-1/2, 1/2]$. We have fixed $[\alpha_1, \alpha_2, \alpha_3, \alpha_4] = [1,2,1/4,1/2]$, as in (\ref{mexweight}).}
\label{bwidplot}
\end{figure}

To illustrate our analysis, we consider the weight (\ref{weight}) given by a sum of modified Bessel functions \cite{owen07}. Using the fact that the corresponding Hankel transform of $K_0(sr)$ is ${\mc H}( \rho , s) = ( \rho^2 + s^2)^{-1}$, we have
\begin{align}
\widehat{w}( \rho ) = \sum_{j=1}^N c_j {\mc H}( \rho , \alpha_j ) .  \label{whatH}
\end{align}
The bump solution (\ref{Ubess}) can then be evaluated by using the formula (\ref{whatH}) along with the identity
\begin{align}
\int_0^{\infty} \frac{J_0( \rho r) J_1( \rho a)}{\rho^2 + s^2} d \rho \equiv {\mc I}(a,r,s) = \left\{ \begin{array}{ll} \frac{1}{s} I_1(s a) K_0(sr) & : r>a, \\ \frac{1}{s^2a} - \frac{1}{s} I_0(s r) K_1(s a) & : r<a, \end{array} \right. \label{intcomp}
\end{align}
where $I_{\nu}$ is the modified Bessel function of the first kind of order $\nu$. We can thus generate an explicit solution for the stationary bump $U(r)$, given
\begin{align*}
U(r) = 2 \pi a \sum_{k=0}^N c_k {\mc I}(a,r, \alpha_k). 
\end{align*}
Applying the threshold condition $U(a) = \kappa$, we can generate an implicit equation relating the bump radius $a$ with the threshold $\kappa$ and weight parameters
\begin{equation} \label{root}
U(a) =  2 \pi a \sum_{k=0}^N \frac{c_k}{\alpha_k}  I_1 (\alpha_k a) K_0( \alpha_k a) = \kappa.
\end{equation}
In general, explicit solutions for $a$ cannot be computed and (\ref{root}) must be solved numerically using root finding algorithms. Note also that satisfaction of the threshold condition (\ref{root}) is not sufficient for proving the existence of a bump. For instance, the possibility of {\em ring} solutions must be eliminated by ensuring there are no other threshold crossing points \cite{owen07,coombes12}.
%In the case of stationary bump solutions, there are usually two solutions corresponding to a stable and unstable fixed point.
Furthermore, we must develop a linear stability analysis to identify those bumps that will persist in the presence of perturbations. This will especially be important in our analysis of the stochastic system (\ref{modeln}), since it will rely on the assumption that the perturbed solutions retains a profile similar to the unperturbed system.

We demonstrate the results of this existence analysis by utilizing a Mexican hat type weight distributions such as \cite{owen07,bressloff11}
\begin{align}
w(r) = c_1 K_0 (r) +c_2 K_0 (2r) +c_3 K_0 (r / 4) + c_4 K_0 (r /2), \label{mexweight}
\end{align}
where we have fixed the spatial scales $[\alpha_1, \alpha_2, \alpha_3, \alpha_4] = [1,2,1/4,1/2]$ and will take $c_1, c_4>0$ and $c_2, c_3<0$ to generate a lateral inhibitory kernel. Typically, weight functions like (\ref{mexweight}) lead to a maximum of two bump solutions, as shown in Fig. \ref{bwidplot}.

%A few examples have been computed in figure (label to be added) for a weight function taking the form \cite{bressloff11}
%$$
%\bigg)
%$$
%where $C,A,\sigma_j$ are selectively chosen coefficients to yield pulse solutions.
% Figure Here
%\begin{figure}[h!]
%\begin{center}
%\includegraphics[scale=0.4]{bifurcation.eps}
%\caption{Example of bifurcation for different choices of $\alpha_4$.}
%\end{center}
%\end{figure}

As mentioned, we must be aware of the possibility of azimuthal instabilities of the stationary bump when developing our linear stability theory \cite{owen07,bressloff13}. Thus, while it may be convenient to analogize the shifting and expanding/contracting perturbations of 1D bumps \cite{amari77} with $D_1$ and circularly symmetric perturbations of 2D bumps \cite{taylor99}, one should be sure not to stop here. A full analysis of azimuthal perturbations to the bump (\ref{polarU}), $D_n$-symmetric perturbations ($n \in \mathbb{Z}$, $n>1$), is necessary since bumps can destabilize through such symmetry-breaking instabilities \cite{owen07,bressloff11}. It is worth noting that this fact was originally identified by Amari in 1978 \cite{amari78}, and other systematic analyses were carried out in the last decade \cite{owen07,coombes12}. 

To determine the stability of stationary solution, we will consider small, smooth perturbations of the stationary bump solution (\ref{polarU}) of the form $u(\x,t) := U(\x)  + \ve \Psi(\x,t)$ where $\ve \ll 1$. We substitute this into equation (\ref{model}) and Taylor expand to linear order to generate the equation
\begin{equation}
\frac{ \partial \Psi (\x,t) } { \partial t} = - \Psi ( \mathbf{x,t}) + \int_{ \mathbb{R}^2 } w(\x - \mathbf{y} ) f'(U(\mathbf{y})) \Psi(\mathbf{y},t) d \mathbf{y}. \label{psiteq}
\end{equation}
Applying seperation of variables $\Psi(\x,t) = \Psi(\x) b(t)$ and rearranging terms results in the solutions $b(t) = e^{\lambda t}$ and 
\begin{equation}
(\lambda + 1)\Psi(\x) = \int_{ \mathbb{R}^2} w(\x - \mathbf{y} ) f'(U(\mathbf{y})) \Psi(\mathbf{y}) d \mathbf{y}.  \label{bheval}
\end{equation}
We can immediately identify neutrally stable perturbations, those corresponding to $\lambda = 0$, by letting $\Psi(\x) = U_{x_j}(\x)$ with $j \in \{1,2 \}$. We apply integration by parts and the definition of a stationary solution $U$ given in (\ref{stationary}) into (\ref{bheval}) to yield
\begin{align}
(\lambda + 1)  U_{x_j}(\x) &=  \int_{ \mathbb{R}^2} w(\x - \mathbf{y} ) f'(U(\mathbf{y})) U_{y_j}(\mathbf{y}) \; d \mathbf{y} = \int_{ \mathbb{R}^2} w(\x - \mathbf{y} ) \frac{ \partial}{\partial y_j} \big( f(U(\mathbf{y})) \big) \; d \mathbf{y}  \nonumber \\
&= \int_{ \mathbb{R}^2} \frac{ \partial}{\partial y_j} \big( w(\x - \mathbf{y} )\big)  f(U(\mathbf{y}))  \; d \mathbf{y}  =  \int_{ \mathbb{R}^2}  \frac{ \partial}{\partial x_j}  \big( w(\x - \mathbf{y} ) \big)  f(U(\mathbf{y}))  \; d \mathbf{y}  \nonumber \\
&=  \frac{ \partial}{\partial x_j} \bigg(  \int_{ \mathbb{R}^2}  w(\x - \mathbf{y} )  f(U(\mathbf{y}))  \; d \mathbf{y}  \bigg) = U_{x_j}(\x). \label{Uprimeeq}
\end{align}
Thus, by the linearity of the integral
\begin{align}
\Psi_Z(\x) = h_1U_{x_1}(\x) + h_2 U_{x_2}(\x) \label{psi0}
\end{align}
generates the class of solutions corresponding to the eigenvalue $\lambda_1 = 0$. Note, we will not have such a class of perturbations in the case of stationary external inputs, as the translation symmetry of (\ref{model}) will then be broken. 

Prior to analyzing other azimuthal perturbations to the bump (\ref{polarU}), we briefly discuss how perturbations of the form (\ref{psi0}) impact the long term bump position. In fact, it is precisely this neutral stability of the bump that leads to purely diffusive motion of the bump in the stochastic model (\ref{modeln}). Specifically, we focus on the case of a Heaviside firing rate function (\ref{Hrate}), so we can track the position $\bd = ( \Delta_1 , \Delta_2) \in \Rset^2$, i.e. the spatial mean, of the bump by utilizing the level set condition $u(\x,t) = \kappa$, which can be written
\begin{align}
\kappa =& u( a + \ve b(\theta,t), \theta,t) = U(a + \ve b(\theta,t)) + \ve \Psi (a + \ve b( \theta, t), \theta, t) \nonumber \\
\kappa  = & U(a) + \ve U'(a) b( \theta , t) + \ve \Psi ( a, \theta , t) + {\mc O}(\ve^2), \label{lsetp}
\end{align}
where $b( \theta , t)$ describes the perturbation of the bump boundary $R(\theta,t) = a + \ve b(\theta, t)$ at angular coordinate $\theta$ and $\ve \ll 1$. Note we can employ the stationary level set condition $U(a) = \kappa$ to cancel leading order terms in (\ref{lsetp}) to yield  \cite{folias05}
\begin{align}
b( \theta ,t) = \frac{\Psi (a,\theta,t)}{|U'(a)|} + {\mc O}(\ve).  \label{bthet}
\end{align}
If we specifically denote $\Psi (a, \theta, t)$ to be a neutrally stable perturbation (\ref{psi0}), then $\Psi (a,\theta,t) = \Psi_Z(a,\theta)$ and $b(\theta, t) = b_1(\theta)$, so that to first order
\begin{align}
b_1(\theta) = \frac{\Psi_Z(a,\theta)}{|U'(a)|} = \frac{h_1 U'(a) \cos \theta + h_2 U'(a) \sin \theta}{|U'(a)|} = - h_1 \cos \theta - h_2 \sin \theta,  \label{b0theta}
\end{align}
where we have computed
\begin{align}
\left. \frac{\pd}{\pd x_1} U(||\x||) \right|_{r=a} = U'(a) \cos \theta, \hspace{1cm} \left. \frac{\pd}{\pd x_2} U(||\x||) \right|_{r=a} = U'(a) \sin \theta.  \label{Udiff}
\end{align}
Thus, the new bump boundary $R_0(\theta) \approx a + \ve b_1(\theta)$ can be approximated in polar coordinates as
\begin{align}
R_0( \theta ) = a  - \ve h_1 \cos \theta - \ve h_2 \sin \theta.  \label{Rthet}
\end{align}
This approximates the long-term perturbation to the bump boundary $\lim_{t \to \infty} R(\theta, t) \approx R_0(\theta)$ by simply using the leading order term in the expansion $b(\theta, t) = b_1(\theta) + \sum_{j=0,j \neq 1}^{\infty} b_j(\theta) \e^{\lambda_j t}$, as $\lim_{t \to \infty} b(\theta,t) = b_1(\theta)$ since all other ${\rm Re}\lambda_j < 0$. This utilizes the well known result that circularly symmetric bump solutions to (\ref{model}) are neutrally stable to $D_1$ symmetric perturbations of the bump boundary \cite{folias05,owen07}.  Now, define the centroid (center of mass) $\bd = (\Delta_1, \Delta_2)$ as the first moments of mass along the $x-$ and $y-$directions scaled by the total mass for the lamina $\Omega_R$ of uniform density enclosed by the curve (\ref{Rthet}). In Cartesian coordinates, these quantities are given by the double integral formulae $\Delta_1 = \int \int_{\Omega_R} x dx dy/ \int \int_{\Omega_R} dx dy$ and $\Delta_2 = \int \int_{\Omega_R} y dx dy/ \int \int_{\Omega_R} dx dy$. By utilizing polar coordinates $(x,y) = (r \cos \theta, r \sin \theta)$, and plugging in the formula for the closed curve $R(\theta)$ (\ref{Rthet}), we can compute:
\begin{align*}
\Delta_1 &= \frac{\int_0^{2 \pi} \int_0^{R( \theta)} r^2 \cos \theta dr d \theta}{ \int_0^{2 \pi} \int_0^{R( \theta )} r dr d \theta} =  \frac{\frac{1}{3} \int_0^{2 \pi} (a - \ve h_1 \cos \theta - \ve h_2 \sin \theta)^3 \cos \theta d \theta}{\frac{1}{2} \int_0^{2 \pi} (a - \ve h_1 \cos \theta - \ve h_2 \sin \theta)^2 d \theta} \\
& =  \frac{\frac{a^2}{3} \int_0^{2 \pi} \left( a - 3 \ve h_1 \cos \theta - 3 \ve h_2 \sin \theta \right)  \cos \theta d \theta}{\frac{a}{2} \int_0^{2 \pi} \left( a - 2 \ve h_1 \cos \theta - 2 \ve h_2 \sin \theta \right) d \theta} + {\mc O}(\ve^2) \approx  - \ve h_1 
\end{align*}
and
\begin{align*}
\Delta_2 &= \frac{\int_0^{2 \pi} \int_0^{R( \theta)} r^2 \sin \theta dr d \theta}{ \int_0^{2 \pi} \int_0^{R( \theta )} r dr d \theta} =  \frac{\frac{1}{3} \int_0^{2 \pi} (a - \ve h_1 \cos \theta - \ve h_2 \sin \theta)^3 \sin \theta d \theta}{\frac{1}{2} \int_0^{2 \pi} (a - \ve h_1 \cos \theta - \ve h_2 \sin \theta)^2 d \theta} \\
& =  \frac{\frac{a^2}{3} \int_0^{2 \pi} \left( a - 3 \ve h_1 \cos \theta - 3 \ve h_2 \sin \theta \right)  \sin \theta d \theta}{\frac{a}{2} \int_0^{2 \pi} \left( a - 2 \ve h_1 \cos \theta - 2 \ve h_2 \sin \theta \right) d \theta} + {\mc O}(\ve^2) \approx  - \ve h_2 .
\end{align*}
Thus a perturbation in the bump profile of the form (\ref{psi0}) will yield a proportional shift in the bump's center of mass $(\Delta_1, \Delta_2) = - \ve(h_1, h_2)$. This foreshadows the impact of noise-induced perturbations to the bump's position, which we explore in section \ref{diffnoise}. Effectively, we will show the primary contribution to the stochastic motion of bumps is the $D_1$-symmetric portion of the spatiotemporal noise, which is filtered by the bump as a spatial translation of its boundary and center of mass.

Lastly, we briefly review the analysis of azimuthal perturbations to the bump boundary. Note that in the case of a Heaviside firing rate function (\ref{Hrate}), then
\begin{align}
f'(U( \x )) = \delta ( U(r) - \kappa ) = \frac{\delta (r-a)}{|U'(a)|},  \label{fderu}
\end{align}
where $U'(a)$ is the normal derivative of $U(\x)$ along the bump boundary $r=a$, in polar coordinates $\rr = (r, \theta)$. Thus, the eigenvalue equation (\ref{bheval}) becomes
\begin{align}
( \lambda + 1) \Psi ( \rr ) = \frac{a}{|U'(a)|} \int_0^{2 \pi} w (|\rr - \ab'|) \Psi (a, \theta' ) d \theta',  \label{Heval}
\end{align}
where $\ab' = (a, \theta')$ in polar coordinates, equivalently $\ab' = a ( \cos \theta, \sin \theta)$ in Cartesian coordinates. Stability of the bump is thus determined by the spectrum of a compact linear operator acting on continuous, bounded functions $\Psi (r, \theta)$ defined on the disc of radius $r \leq a$. The essential spectrum contains functions $\Psi ( \rr )$ that vanish on the boundary $\Psi (a, \theta ) = 0$ for all $\theta$, so $\lambda = -1$, contributing to no instabilities. We can thus identify the discrete spectrum by setting $\rr = \ab = (a, \theta )$ in (\ref{Heval}), so
\begin{align*}
( \lambda + 1) \Psi (a, \theta ) = \frac{a}{|U'(a)|} \int_0^{2 \pi} w \left( 2 a \sin \left( \frac{\theta - \theta'}{2} \right) \right) \Psi(a, \theta' ) d \theta',
\end{align*} 
where we have utilized the identity
\begin{align*}
| \ab - \ab' | = \sqrt{2a^2 - 2a^2 \cos ( \theta - \theta')} = 2 a \sin \left( \frac{\theta - \theta'}{2} \right).
\end{align*}
Any perturbation can be decomposed into the infinite series of Fourier modes $\Psi (a, \theta) = \sum_{n = 0}^{\infty} A_n \Psi_n (\theta) + B_n \bar{\Psi_n} ( \theta )$, where $\Psi_n ( \theta ) = \e^{i n \theta}$ and $\bar{\Psi}_n( \theta) = \e^{- in \theta}$ \cite{folias04,owen07}. We can thus compute all the eigenvalues of the discrete spectrum by evaluating the expression
\begin{align*}
\lambda_n = - 1 + \frac{a}{|U'(a)|} \int_0^{2 \pi} w(2a \sin ( \theta/ 2)) \e^{- in \theta}
\end{align*}
each associated with $\Psi_n( \theta )$ for $n \in \mathbb{Z}_{\geq 0}$. Note, $\lambda_n$ will always be real, since rescaling $\theta \mapsto 2 \theta$:
\begin{align*}
{\rm Im}\{ \lambda_n \} = - \frac{2 a}{|U'(a)|} \int_0^{\pi} w(2 a \sin \theta ) \sin (2n \theta ) d \theta = 0.
\end{align*}
Therefore, the eigenvalue is real
\begin{align}
\lambda_n = {\rm Re}\{ \lambda_n \} = -1 + \frac{a}{|U'(a)|} \int_0^{2 \pi} w(2a \sin (\theta/2)) \cos (n \theta) d \theta.  \label{azieval}
\end{align}
Note, the bump profile perturbation $\Psi ( \rr, t)$ will be related to the bump boundary perturbation $b( \theta,t)$ via the formula (\ref{bthet}), as discussed above. The $n$th order boundary perturbation has $D_n$ symmetry; e.g. $n=0$ uniformly expands/contracts the bump, $n=1$ shifts the bump.

By specifying the weight function $w(r)$, we can compute the eigenvalue (\ref{azieval}) explicitly using Bessel functions to evaluate the integral
\begin{align*}
\int_0^{2 \pi} w(| \ab - \ab' | ) \cos (n \theta') d \theta' &= \int_0^{2 \pi} \left( \int_0^{\infty} \widehat{w}(\rho) J_0 (\rho |\ab - \ab'|) \rho d \rho \right) \cos \theta' d \theta' \\ &= 2 \pi \int_0^{\infty} \widehat{w} ( \rho ) J_n ( \rho r) J_n ( \rho a ) \rho d \rho.
\end{align*}
Thus, we can write (\ref{azieval}) as
\begin{align*}
\lambda_n = - 1 + \frac{\int_0^{\infty} \widehat{w}( \rho ) J_n ( \rho r) J_n ( \rho a) \rho d \rho}{\int_0^{\infty} \widehat{w}( \rho ) J_1 ( \rho r ) J_1 ( \rho a ) \rho d \rho}.
\end{align*}
Since we know $\lambda_1 = 0$, the bump will be stable if $\lambda_n < 0$ for all $n \neq 1$. Employing the general weight distribution (\ref{weight}), we find
\begin{align}
\lambda_n = - 1 + \frac{\sum_{j=1}^N c_j K_n ( \alpha_j a) I_n ( \alpha_j a)}{\sum_{j=1}^N c_j K_1( \alpha_j a) I_1( \alpha_j a)}.  \label{lamnbess}
\end{align}
More specifically, we could focus on the Mexican hat weight distribution (\ref{mexweight}) along with the parameters given in Fig. \ref{bwidplot}{\bf A}. Indeed, checking the formula for the eigenvalues (\ref{lamnbess}), we find $\lambda_n < 0$ ($n \neq 1$) for all solutions along the upper branch of wide bump solutions.

\subsection{Effective equations for stochastic bump motion} \label{diffnoise}
We now explore how noise impacts the long term position of bumps in the network (\ref{modeln}). Previous authors have analyzed the impact of noise on waves in reaction-diffusion \cite{panja04,armero98,sagues07} and neural field models \cite{brackley07,bressloff12,kilpatrick13} using small-noise expansions, but those studies tend to be on one-dimensional domains. Noise causes waves to execute an effective Brownian motion in their instantaneous position. As we will show, this analysis naturally extends to the effective stochastic dynamics of bumps in two-dimensional (2D) domains. The position $\bd (t) = ( \Delta_1(t), \Delta_2(t))$ wanders diffusively, as a 2D random walk, as long as noise is small so that the profile of the bump remains close to the solution of the deterministic system (\ref{model}). As mentioned, this relies upon the neutral stability of the noise free system; different behavior will arise when we break this symmetry with inputs, shown in section \ref{stimpin}.

\begin{figure}[tb]
\begin{center} \includegraphics[width=13cm]{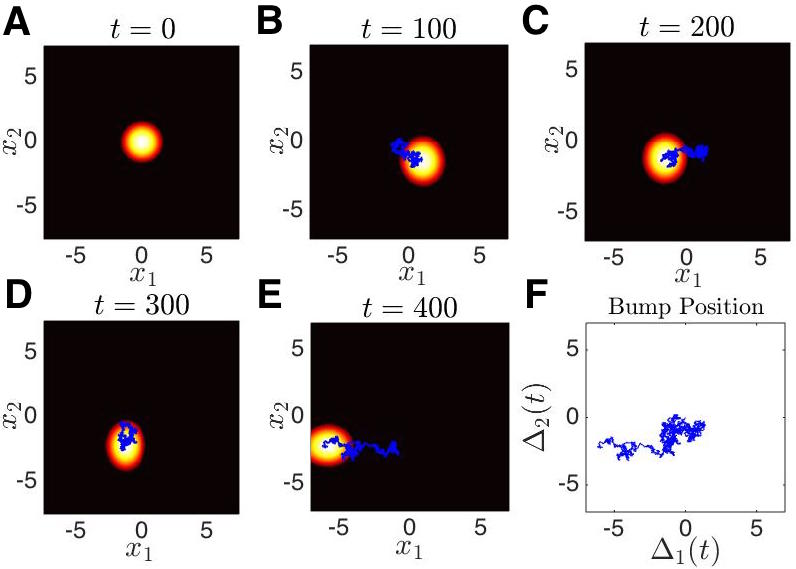} \end{center}
\caption{Numerical simulation of the stochastic neural field (\ref{modeln}) on the plane $\Rset^2$ with Heaviside firing rate (\ref{Hrate}) and Bessel function weight (\ref{weight}). ({\bf A}-{\bf E}) Snapshots of a simulation of a bump wandering on the plane, due to noise with spatial correlation function $C(\x) = \cos (\x)$, at time points $t=0,100,200,300,400$. Thin lines represents the stochastic trajectory of the bump during the time between the previous and current snapshot. ({\bf F}) Plot of the trajectory of the bump centroid for $t \in [0,400]$ demonstrates how its stochastic trajectory behaves as 2D Brownian motion. Other parameters are $\kappa=0.2$, $\epsilon = 0.04$, and $w$ is (\ref{mexweight}) with $[c_1, c_2, c_3, c_4 ]=[5/3, -5/3, -1/2, 1/2]$. We approximate the center of mass of the bump using $\argmax_{\x} u(\x,t)$.}
\label{pdiffsim}
\end{figure}

To begin, we assume that the weak additive noise (of ${\mc O}(\ve^{1/2})$) in (\ref{modeln}) affects the bump in two ways. Both are weak (${\mc O}(\ve^{1/2})$) compared to the amplitude of the bump, allowing us to exploit regular perturbation theory to analyze the Langevin equation (\ref{modeln}). First, the bump diffuses from its original position $x$ on long timescales according to the stochastic variable $\bd (t) = ( \Delta_1(t), \Delta_2(t))$ (see Fig. \ref{pdiffsim}, for example). Second, there are fluctuations in the bump profile on short timescales \cite{armero98}, according to the expansion $\ve^{1/2} \Phi + \ve \Phi_1 + \ve^{3/2} \Phi_2 + \cdots$. This suggests the following ansatz for the impact of noise on the bump solution $U( \x )$:
\begin{equation} \label{uanz}
u(\x,t) = U(\x - \bd(t) ) + \ve^{1/2} \Phi (\x - \bd(t),t) + \cdots.
\end{equation}
Substituting (\ref{uanz}) into (\ref{modeln}) and truncating to ${\mc O}(1)$, we find that $U(\x)$ still satisfies (\ref{stationary}). Proceeding to linear order in $\ve^{1/2}$, we find
\begin{equation} \label{perteq}
d\Phi(\x,t) = {\mc L} \Phi(\x,t) + \ve^{-1/2} \nabla U(\x) \cdot d \bd(t) + dW(\x,t),
\end{equation}
where $\nabla U (\x) = (U_{x_1}(\x), U_{x_2}(\x))^T$ denotes the gradient of $U(\x)$ and ${\mc L}$ is a non-self-adjoint linear operator of the form 
\begin{align}
{\mc L} p(\x) = - p(\x) + \int_{\Rset^2} w(\x-\y) f'(U(\y)) p(\y) d \y  \label{linop}
\end{align}
for any $L^2$ integrable function $p( \x ) $ on $\Rset^2$. We ensure a bounded solution to (\ref{perteq}) exists by requiring the inhomogeneous part is orthogonal to all elements of the null space of the adjoint operator ${\mc L}^*$. The adjoint is defined by utilizing the $L^2$ inner product
\begin{align*}
\int_{\Rset^2} \left[ {\mc L}p( \x ) \right] q( \x ) d \x = \int_{\Rset^2} p(\x) \left[ {\mc L}^* q( \x ) \right] d \x,
\end{align*}
where $p( \x )$, $q( \x )$ are $L^2$ integrable on $\Rset^2$. Thus
\begin{align}
{\mc L}^* q( \x ) = -q ( \x ) + f'( U( \x )) \int_{\Rset^2} w( \x - \y ) q ( \y ) d \y. \label{adjop}
\end{align}
The span of the nullspace of ${\mc L}^*$ can be described by two functions, which we can compute explicitly for a general firing rate function. That is, we can find an infinite number of solutions to the null space equation
\begin{align}
\varphi(\x) = f'(U(\x))\int_{\Rset^2} w(\x-\y) \varphi (\y) d \y,  \label{nuleq}
\end{align}
which can always be decomposed into a linear combination of two functions $\varphi_1(\x)$ and $\varphi_2(\x)$. Specifically, we take $\varphi_1( \x ) = f'(U(\x)) U_{x_1}( \x )$ and $\varphi_2 ( \x) = f'(U(\x)) U_{x_2} ( \x)$, and note that by plugging into (\ref{nuleq}), we have
\begin{align*}
f'(U(\x)) U_{x_j}(\x) = f'(U(\x)) \int_{\Rset^2} w( \x - \y) f'(U(\x)) U_{x_j}(\y) d \y,
\end{align*}
which holds due to to the equations in (\ref{Uprimeeq}). Thus, we can derive an effective equation for the position variable $\bd (t)$ by taking the inner product of $\varphi_1$ and $\varphi_2$ with both sides of (\ref{perteq}) to yield
\begin{align} \label{effeq}
\int_{ \Rset^2} f'(U(\x)) U_{x_1}(\x) \left(  U_{x_1}(\x) d \Delta_1 (t) + U_{x_2}(\x) d \Delta_2 (t) + \ve^{1/2} dW(\x,t) \right) d \x &= 0  \\
\int_{ \Rset^2} f'(U(\x)) U_{x_2}(\x) \left(  U_{x_1}(\x) d \Delta_1 (t) + U_{x_2}(\x) d \Delta_2(t) + \ve^{1/2} dW(\x,t) \right) d \x &= 0. \nonumber
\end{align}
Moreover, we can exploit the odd and even symmetries of the spatial derivatives $U_{x_1}$ and $U_{x_2}$ that must hold since $U(\x)$ is radially symmetric. Namely, $U_{x_1}$ is odd-symmetric along the $x_1$-axis and even along the $x_2$-axis, and $U_{x_2}$ is even-symmetric along the $x_1$-axis and odd along the $x_2$-axis. Lastly, $f'(U(\x))$ is radially symmetric since $U(\x)$ is. This means $\int_{\Rset^2} f'(U(\x)) U_{x_1}(\x) U_{x_2}(\x) d \x = 0$. This allows us to rearrange the system (\ref{effeq}), yielding a pair of independent equations for the diffusion of the bump along the $x_1$ and $x_2$ axes
\begin{equation} \label{effeq2}
d \Delta_j (t) = - \ve^{1/2} \frac{\int_{\Rset^2} f'(U(\x)) U_{x_j}(\x) d W(\x,t) d \x}{ \int_{\Rset^2} f'(U(\x)) U_{x_j}^2(\x) d \x }, \; \; \; \; j \in \{ 1,2 \} . 
\end{equation}
With the stochastic system (\ref{effeq2}) in hand, we can approximate the effective diffusivity of the bump. First, note that the mean position of the bump averaged across realizations does not change in time ($\langle \bd (t) \rangle = (0,0)^T$) since noise has mean zero ($\langle W(\x, t) \rangle = 0$). Computing the variance of the stochastic variable $\bd(t)$, we find it obeys pure diffusion in two-dimensions:
\begin{align}
\langle \Delta_j(t)^2 \rangle &= \ve \frac{\int_{\Rset^2} \int_{\Rset^2} f'(U(\x)) U_{x_j}(\x) f'(U(\y)) U_{x_j}(\y) \langle W( \x , t) W ( \y , t) \rangle d \y d \x}{\left[  \int_{\Rset^2} f'(U(\x)) U_{x_j}^2(\x) d \x \right]^2} \nonumber \\
\langle \Delta_j(t)^2 \rangle &=  \ve D_j t,   \; \; \; \; j \in \{ 1,2 \}, \label{deljvar}
\end{align}
and using the definition of the spatiotemporal noise $W( \x ,t)$ in (\ref{noise}), we find
\begin{align}
D_j= \frac{\int_{\Rset^2} \int_{\Rset^2} f'(U(\x)) U_{x_j}(\x) f'(U(\y)) U_{x_j}(\y) C( \x - \y)  d \y d \x}{\left[  \int_{\Rset^2} f'(U(\x)) U_{x_j}^2(\x) d \x \right]^2}.  \label{effdiff}
\end{align}
This allows us to calculate the effective diffusion of bumps in the stochastic planar neural field (\ref{modeln}). We simply need to compute the constituent functions $U_{x_j}$ and $f'(U)$ and evaluate the integrals in (\ref{effdiff}), which we now do in the case of Heaviside firing rates (\ref{Hrate}).

\begin{figure}[tb]
\begin{center} \includegraphics[width=6cm]{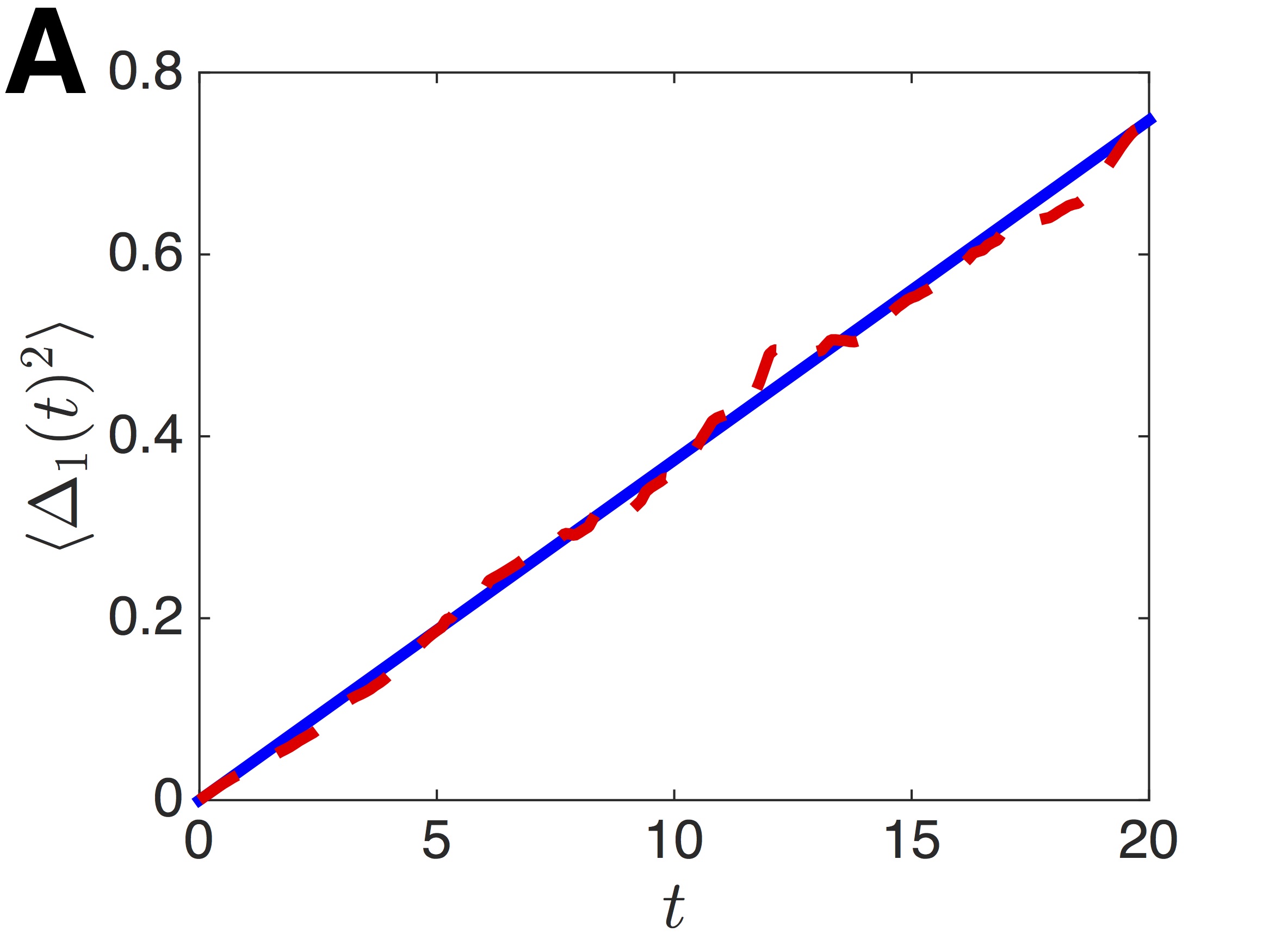} \includegraphics[width=6cm]{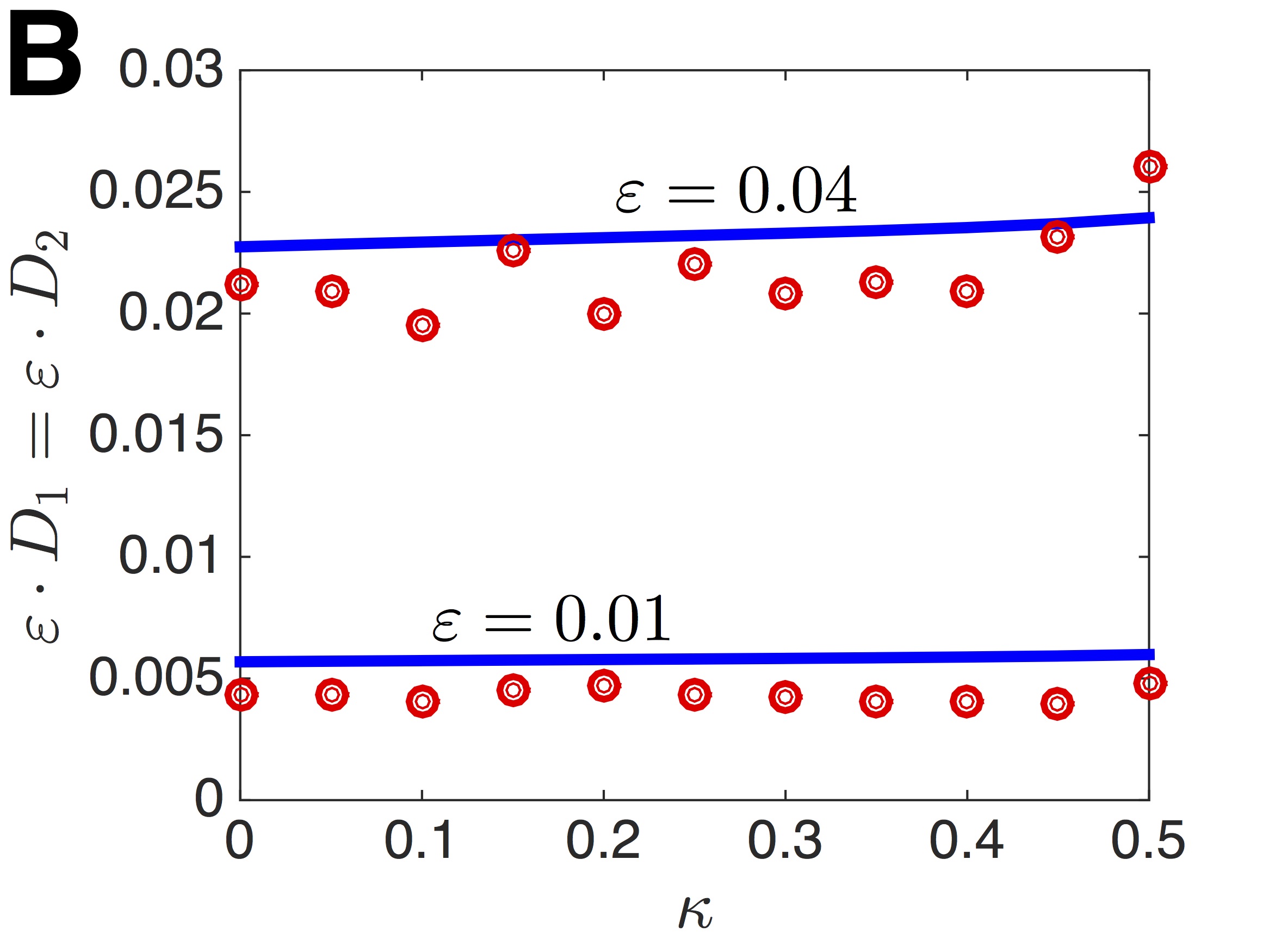} \end{center}
\caption{Variance of the bump position $\bd (t) = ( \Delta_1(t), \Delta_2 (t))$ evolving according to the stochastic model (\ref{modeln}) with Heaviside firing rate (\ref{Hrate}) and Bessel function weight (\ref{weight}). ({\bf A}) Long-term variance $\langle \Delta_1^2 \rangle \rangle$ scales linearly as pure diffusion, and the slope is given by the diffusion coefficient $D$ defined by the formula (\ref{diffjH}). Parameters are $\kappa=0.2$ and $\ve = 0.05$. ({\bf B}) Plot of the effective diffusion coefficient $\ve \cdot D$ versus the activity threshold $\kappa$. The weight function $w$ is (\ref{mexweight}) with $[c_1, c_2, c_3, c_4 ]=[5/3, -5/3, -1/2, 1/2]$. Numerical variances are computed using 1000 realizations each.}
\label{pdiffcoef}
\end{figure}

\subsection{Explicit results for the Heaviside firing rate}
We now show that we can explicitly calculate the effective diffusion coefficient (\ref{effdiff}) in the case of a Heaviside firing rate function (\ref{Hrate}), weight kernel comprised of modified Bessel functions (\ref{weight}), and the following cosine noise correlation function derived in (\ref{derivecc}):
\begin{equation}
C(\x) = \cos(x_1 ) + \cos(x_2).  \label{coscorr}
\end{equation}
Analogous to our linear stability calculations, by selecting a Heaviside firing rate function (\ref{Hrate}), we find that the associated functional derivative $f'(U)$ is given by (\ref{fderu}). Thus, the domain of integration of the terms in (\ref{effdiff}) collapse from $\Rset^2$ to the closed curve $r = a$, in polar coordinates $\rr = (r, \theta)$. Furthermore, the spatial derivatives $U_{x_1}$ and $U_{x_2}$ are given by the formulas (\ref{Udiff}), so that we can simply rewrite the term in the denominator of (\ref{effdiff}) as an integral over the angular coordinate $\theta$
\begin{align*}
 \left( \int_{ \Rset^2 } f'(U(\x)) U_{x_j}^2 (\x)  d \x \right)^2  =& \left( \frac{1}{|U'(a)|} \int_0^{2 \pi} \int_0^{\infty} \delta(r-a) \left[ U'(a) \cos \theta \right]^2 r d r d \theta \right)^2 \\
 =& \left( \frac{a}{|U'(a)|} \int_0^{2 \pi}  \left[ U'(a) \cos \theta \right]^2 d \theta \right)^2 = a^2 \pi^2 U'(a)^2.
 \end{align*}
We can apply a similar approach to the calculation of the numerator of (\ref{effdiff}), given by
\begin{align*}
a^2 \pi^2 U'(a)^2 D_j&= \int_{\Rset^2} \int_{\Rset^2} f'(U(\x)) U_{x_j}(\x) f'(U(\y)) U_{x_j}(\y) C( \x - \y)  d \y d \x \\ 
&= a^2 \int_0^{2 \pi} \int_0^{2 \pi} \cos \theta \cos \phi \ C(a (\cos \theta, \sin \theta)^T - a ( \cos \phi, \sin \phi)^T) d \theta d \phi.
\end{align*}
Rewriting by using the cosine correlation function (\ref{coscorr}) and utilizing the identity $\cos (x-y) = \cos x \cos y + \sin x \sin y$, we have
\begin{align*}
\pi^2 U'(a)^2 D_j =& \left( \int_0^{2 \pi} \cos( a \cos \theta ) \cos \theta d \theta \right)^2 + \left( \int_0^{2 \pi} \cos( a \sin \theta) \cos \theta d \theta \right)^2 \\
 &+ \left( \int_0^{2 \pi}  \sin( a \cos \theta) \cos \theta d \theta \right)^2 + \left( \int_0^{2 \pi} \sin( a \sin \theta) \cos \theta d \theta \right)^2.
\end{align*}
 Applying the substitution $v = a \sin \theta$, we find that
 \begin{align*}
 \int_0^{2 \pi} \cos( a \sin \theta) \cos \theta d \theta = \frac{1}{a} \int_{-a}^{a} \cos v d v + \frac{1}{a} \int_{a}^{-a} \cos v d v =& \ 0, \\
  \int_0^{2 \pi} \sin( a \sin \theta) \cos \theta d \theta = \frac{1}{a} \int_{-a}^{a} \sin v d v + \frac{1}{a} \int_{a}^{-a} \sin v d v =& \ 0.
 \end{align*}
 Furthermore, breaking up the domain of integration of the first summand, we find
 \begin{align*}
 \int_0^{2 \pi} \cos ( a \cos \theta ) \cos \theta d \theta =& \int_{- \pi/2}^{\pi/2} \cos (a \cos \theta ) \cos \theta d \theta + \int_{\pi/2}^{3\pi/2} \cos (a \cos \theta ) \cos \theta d \theta \\
 =& \int_{- \pi/2}^{\pi/2} \cos (a \cos \theta ) \cos \theta d \theta + \int_{-\pi/2}^{\pi/2} \cos (a \cos (\theta + \pi) ) \cos (\theta + \pi) d \theta \\
 =& \int_{- \pi/2}^{\pi/2} \cos (a \cos \theta ) \cos \theta d \theta - \int_{-\pi/2}^{\pi/2} \cos (a \cos \theta ) \cos \theta d \theta = 0.
 \end{align*}
 Lastly, we must compute the remaining summand, for which we make use of integration by parts
 \begin{align*}
 \int_0^{2 \pi} \sin ( \cos \theta ) \cos \theta d \theta = 2 a \int_0^{\pi} \sin^2 \theta \cos ( a \cos \theta ) d \theta = 2 \pi J_1( a),
 \end{align*}
where we have made use of the explicit integral representation of a Bessel function of the first kind of order $\nu$:
\begin{align*}
J_{\nu} (z) = \frac{1}{\pi} \frac{z^{\nu}}{(2 \nu -1)!!} \int_0^{\pi} \sin^{2 \nu} \theta \cos ( z \cos \theta ) d \theta.
\end{align*}
Thus, we can finally write
\begin{align}
D_j = \frac{4 J_1(a)^2}{ U'(a)^2}, \ \ \ \ j=1,2.  \label{diffjH}
\end{align}
Effective diffusion along both the $x_1$ and $x_2$ axes is identical, due to radial symmetry of the bump along with the $D_4$ symmetry of the correlation function (\ref{coscorr}). We demonstrate our analytical results, in comparison to statistics from numerical simulations, in Fig. \ref{pdiffcoef}.
 
\section{Stimulus-pinned bumps in $\Rset^2$}  \label{stimpin}
In this section, we explore how the interaction of external inputs and noise determines the stochastic dynamics of bumps. In previous work, we have shown in one-dimensional domains that both external inputs \cite{kilpatrick13} and coupling between bumps in multiple layers \cite{kilpatrick13b,kilpatrick15,bressloff15} can help stabilize bumps to the translating perturbations of noise. Inputs pin bumps in place so their motion is mostly restricted to the peak(s) of the input function, and the stochastic variable describing the bump's location can be approximated with a mean-reverting (Ornstein-Uhlenbeck) process. Thus, we consider an external stationary stimulus $I(\x)$ acting on our stochastic system (\ref{modeln}), and our modified model takes the form
\begin{equation}
du(\x,t) = \left( - u(\x,t) + \int_{ \Rset^2} w(\x-\y) f( u(\x,t) ) \; d\y + I(\x) \right) dt  + \ve^{1/2} dW(\x,t).  \label{inplang}
\end{equation}
The primary forms of input we employ are a radially symmetric Gaussian
\begin{align}
I(\x) = I(r) = A_0 \e^{ -r^2/ \sigma^2}, \label{radgauss}
\end{align}
and a translationally symmetric Gaussian
\begin{align}
I(\x) = I(x_1) = A_0 \e^{ -x_1^2/ \sigma^2}, \label{transgauss}
\end{align}
where $I(x_1)$ denotes independence from the second coordinate of the spatial vector $\x = (x_1,x_2)$. We begin by reviewing the existence and stability of radially symmetric bumps in the noise free system ($\ve \to 0$), as this foreshadows the impact of inputs on the stochastic dynamics of bumps. Essentially, the local stability of bumps to translating perturbations is altered by the input's spatial heterogeneity.

\subsection{Existence and stability of bumps}
We begin by constructing the modified equations for radially symmetric bump solutions $u(\x,t) = U(\x) = U(r)$ to the model (\ref{inplang}) in the absence of noise ($\ve \to 0$). Previous work has shown that inputs can stabilize stationary bumps in purely excitatory neural field models that incorporate linear adaptation \cite{folias04} by altering the evolution of expanding/contracting ($O(2)$-symmetric) perturbations to bump profiles. Here, we show inputs stabilize bumps to translating ($D_1$-symmetric) perturbations. We focus on the case where the external input $I(\x)$ is rotationally symmetric, so in polar coordinates, $I(r,\theta) = I(r,\theta + s)$ for $s \in [0,2 \pi]$. Thus, our stationary bump solution satisfies the stationary equation
\begin{equation}
 U(||\x||) = \int_{\Rset^2} w(\x - \y) f(U(\y)) d \y + I(\x),
\end{equation}
and by changing to polar coordinates $\x = (x_1,x_2) \mapsto (r, \theta)$, we have
\begin{align}
U(r) = \int_0^{2 \pi} \int_0^{\infty} w(|\rr - \rr'|) f(U(\rr')) r' d r' d \theta' + I(r).  \label{inpsol1}
\end{align}
Assuming a Heaviside firing rate function (\ref{Hrate}), the integral in (\ref{inpsol1}) collapses to a compact domain
\begin{align}
U(r) = \int_0^{2 \pi} \int_0^a w(|\rr - \rr'|) r' d r' d \theta' + I(r),  \label{inpsol2}
\end{align}
where $r \equiv a$ defines the boundary bump as in the input-free case (\ref{polhu}). We can evaluate (\ref{inpsol2}) explicitly by assuming the weight function formula is a sum of modified Bessel functions (\ref{weight}), finding
\begin{align*}
U(r) = 2 \pi a \sum_{j=1}^N c_k {\mc I}(a,r, \alpha_k) + I(r),
\end{align*}
where ${\mc I}(a,r,s)$ is defined by the formula (\ref{intcomp}). To relate the bump radius $a$ to the threshold $\kappa$ of the Heaviside firing rate function (\ref{Hrate}), we apply the condition $U(a) = \kappa$, which can be written
\begin{align*}
U(a) = 2 \pi a \sum_{j=1}^N \frac{c_k}{\alpha_k} I_1(\alpha_k a) K_0( \alpha_k a) + I(a) = \kappa.
\end{align*}
One can then solve this nonlinear equation using numerical root finding.

To determine the stability of the input-driven stationary bump solution (\ref{inpsol1}), we study the impact of small smooth pertubations by employing the ansatz $u(\x,t) = U(\x) + \ve \Psi(\x,t)$ where $\ve \ll 1$. Substituting this expansion into (\ref{inplang}), when $W \equiv 0$, and truncating to linear order yields the equation (\ref{psiteq}) as in the input-free case. The main difference is that bumps are defined by (\ref{inpsol1}), incorporating the input term $I( \x )$. Furthermore, applying separation of variables $\Psi ( \x, t) = \Psi ( \x ) \e^{\lambda t}$ and rearranging terms yields the eigenvalue equation (\ref{bheval}). However, when $U(\x)$ satisfies (\ref{inpsol1}), such bumps are no longer neutrally stable to perturbations that shift their position. It is straightforward to show this in the case of a Heaviside firing rate function (\ref{Hrate}), in which case eigenvalues associated with the Fourier modes $\Psi_n ( \theta) = \e^{i n \theta}$ are given by the expression (\ref{azieval}), so the eigenvalue $\lambda_1$ associated with shifts perturbations $\Psi_1( \theta) = \e^{i \theta}$ is given by the formula
\begin{align}
\lambda_1 = - 1 + \frac{a}{|U'(a)|} \int_0^{2 \pi} w(2 a \sin ( \theta/2)) \cos ( \theta )  d \theta,  \label{lam1inp}
\end{align}
and since $U(r)$ is given by (\ref{inpsol2}), then
\begin{align*}
|U'(a)| = a \int_0^{2 \pi} w(2a \sin ( \theta/2)) \cos ( \theta ) d \theta  - I'(a).
\end{align*}
Thus, we can rewrite (\ref{lam1inp}) as
\begin{align*}
\lambda_1 = \frac{I'(a)}{a \int_0^{2 \pi} w(2a \sin ( \theta/2)) \cos ( \theta ) d \theta  - I'(a)} < 0,
\end{align*}
where the inequality holds when $I(r)$ is a monotone decreasing function, as we have assumed for the radially symmetric Gaussian (\ref{radgauss}). This indicates the bump will be linearly stable to perturbations that alter its position, indicative of the mean-reverting stochastic dynamics that emerge when weak noise is considered in (\ref{inplang}). Eigenvalues of all Fourier modes, representing azimuthal perturbations of the bump (\ref{inpsol2}), can be computed numerically from the formula (\ref{azieval}).

% and application of seperation of variables $\Psi(\x,t) = \Psi(\x)b(t)$ as before, our solution again has form $b(t) = e^{\lambda t}$ and
%\begin{equation}
%(\lambda + 1)\Psi(\x) = \int_{ \mathbb{R}^2} w(\x - \mathbf{y} ) f'(U(\mathbf{y})) \Psi(\mathbf{y}) d \mathbf{y}
%\end{equation}
%However, note that $\Psi (\x ) := U_{x_j} (\x)$ may no longer give $\lambda_0 = 0$ as the translation symmetry has been broken. To ascertain the stability, we look for solutions of the form $e^{in\theta}$ as in Section 3.1 and find $\lambda_n$ satisfies
%\begin{equation}
%\lambda_n = - 1 + \frac{\sum_{j=1}^N c_j K_n ( \alpha_j a) I_n ( \alpha_j a)}{\sum_{j=1}^N c_j K_1( \alpha_j a) I_1( \alpha_j a)}. 
%\end{equation}
%The eigenvalue $\lambda_1 = 0$ will always hold for a radially symmetry solution, independent of input. This proves the bump is marginally stable with respect to rotations. 
%\begin{equation} \label{weight_ex}
%w(r) = K_0( r ) - K_0 (2r) - A( K_0(a / \sigma ) - K_0 (2a / \sigma )
%\end{equation}
%\zcomment{Rather than stating this, show the details of deriving this inequality.}
%From (paper), it is enough to show
%$$
%2 a \int_0^{2 \pi} w( 2 a \sin \theta ) d \theta < |U'(a)|
%$$
%where $a$ is the radius of the stationary pulse solution. By using the identity
%$$
%\int_0^\pi K_0( 2 a \sin \theta ) d \theta = \pi I_0(a) K_0 (a)
%$$
%we only need to show
%$$
%\sum_{k=1}^4 c_k  K_0( \alpha_k a) I_0(\alpha_k a) - \bigg| \sum_{k=1}^4 c_k K_1(\alpha_k a) I_1 (\alpha_k a) \bigg| < 0
%$$
%This can be verified numerically. \\ \\

\begin{figure}[tb]
\begin{center} \includegraphics[width=13cm]{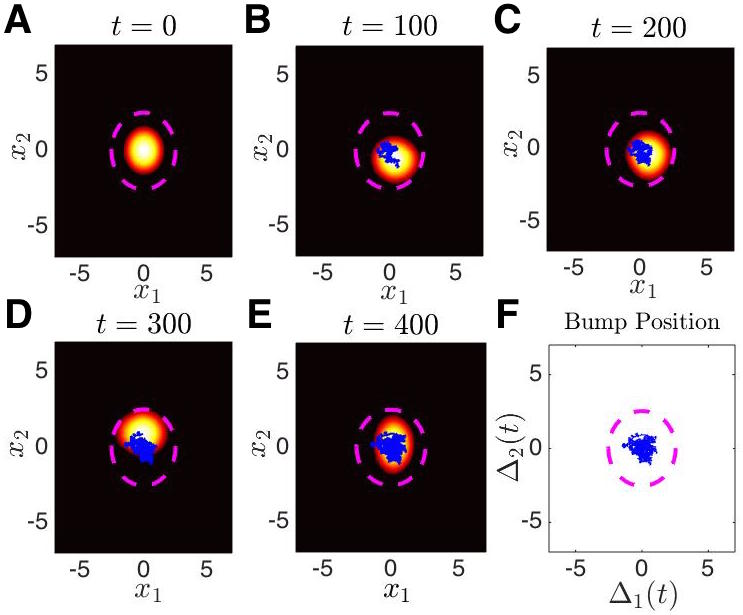} \end{center}
\caption{Numerical simulation of the stochastic neural field (\ref{inplang}) on the plane $\Rset^2$ with Heaviside firing rate (\ref{Hrate}) and Bessel function weight (\ref{weight}), subject to radially symmetric Gaussian input (\ref{radgauss}). Thin lines represents the stochastic trajectory of the bump during the time between the previous and current snapshot. Dashed circle is a plot of the level set $I(\x) = \kappa$. ({\bf A}-{\bf E}) Snapshots of a simulation of a bump wandering on the plane, due to noise with spatial correlation function $C(\x) = \cos (\x)$, at time points $t=0,100,200,300,400$. Input causes the trajectory to stay in the vicinity of the peak of the radially symmetric Gaussian (\ref{radgauss}) at the origin $(x_1,x_2) = (0,0)$. ({\bf F}) Plot of the trajectory of the bump centroid for $t \in [0,400]$ demonstrates how its stochastic trajectory behaves as 2D OU process. Parameters are $\ve = 0.04$, $\kappa = 0.2$, $A_0 = 1$, $\sigma = 2$, and $w$ is (\ref{mexweight}) with $[c_1, c_2, c_3, c_4 ]=[5/3, -5/3, -1/2, 1/2]$.}
\label{fig4}
\end{figure}

\begin{figure}[tb]
\begin{center} \includegraphics[width=13cm]{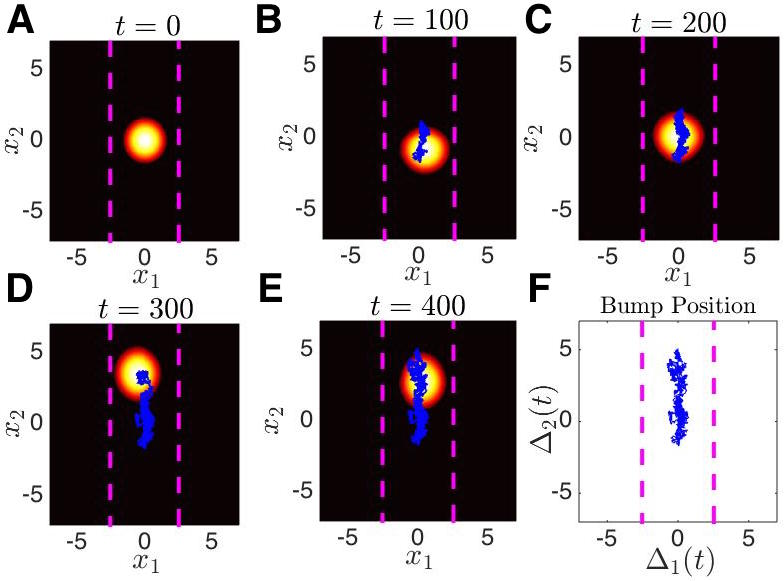} \end{center}
\caption{Numerical simulation of the stochastic neural field (\ref{inplang}) on the plane $\Rset^2$ with Heaviside firing rate (\ref{Hrate}) and Bessel function weight (\ref{weight}), subject to translationally symmetric Gaussian input (\ref{transgauss}). Thin lines represents the stochastic trajectory of the bump during the time between the previous and current snapshot. Dashed lines ar a plot of the level set $I(\x) = \kappa$. ({\bf A}-{\bf E}) Snapshots of a simulation of a bump wandering on the plane, due to noise with spatial correlation function $C(\x) = \cos (\x)$, at time points $t=0,100,200,300,400$. Input cause the trajectory to stay in the vicinity of the peak of the translationally symmetric Gaussian (\ref{transgauss}) along the line $x_1 = 0$. ({\bf F}) Plot of the trajectory of the bump centroid for $t \in [0,400]$ demonstrates how its stochastic trajectory behaves as 2D OU process. Parameters are $\ve = 0.04$, $\kappa = 0.2$, $A_0 = 1$, $\sigma = 2$, and $w$ is (\ref{mexweight}) with $[c_1, c_2, c_3, c_4 ]=[5/3, -5/3, -1/2, 1/2]$.}
\label{fig5}
\end{figure}

\subsection{Stochastic bump motion in the presence of weak inputs} Our analysis of the stochastic motion of bumps in the stationary input-driven system (\ref{inplang}) employs a similar approach to our analysis of the input-free ($I(\x) \equiv 0$) system (\ref{modeln}). However, the effective equations that emerge are no longer translationally invariant, now depending on the spatial heterogeneity imposed by the input $I( \x )$. Our analysis assumes that inputs are weak, having the same amplitude as the noise term \cite{bressloff15}, so we write $I( \x ) = \ve^{1/2} \tilde{I}(\x)$. In this case, the ${\mc O}(1)$ terms are identical to the input-free deterministic system (\ref{model}) with stationary bump solution (\ref{stationary}). We can then derive a system of nonlinear stochastic differential equations for the effective motion of the bump's position $\bd (t) = (\Delta_1(t), \Delta_2(t))$, which we can then truncate to a multivariate Ornstein-Uhlenbeck (OU) process assuming $\bd (t)$ remains small. This captures the fact that bumps are systematically drawn back to the location of the peak(s) of the external input as demonstrated in Fig. \ref{fig4} and Fig. \ref{fig5}.

Employing the ansatz (\ref{uanz}), assuming the bump's profile has fluctuations $\ve^{1/2} \Phi (\x,t)$ on fast timescales and stochastically varying position $ \bd (t)$ on longer timescales, we truncate to linear order in $\ve^{1/2}$ and find
\begin{equation}
d \Phi(\x,t) = \mathcal{L} \Phi(\x,t) +  \ve^{-1/2} \nabla U(\x) \cdot d \mathbf{\Delta}(t)  + dW(\x,t) + \tilde{I}(\x + \mathbf{\Delta}(t)),  \label{phiinp}
\end{equation}
where $\nabla U( \x ) = ( U_{x_1}(\x), U_{x_2}(\x))^T$ and ${\mc L}$ is the non-self-adjoint linear operator (\ref{linop}) with adjoint ${\mc L}^*$ given by (\ref{adjop}). As before, the nullspace of ${\mc L}^*$ is spanned by the two functions $\varphi_1( \x) = f'(U(\x)) U_{x_1}(\x)$ and $\varphi_2(\x) = f'(U(\x)) U_{x_2}(\x)$. Thus, we can enforce solvability of the ${\mc O}(\ve^{1/2})$ equation (\ref{phiinp}) to yield the pair of nonlinear stochastic differential equations
\begin{align}
d \Delta_j (t) = - \ve^{1/2} G_j( \bd (t) ) d t - \ve^{1/2} d {\mc W}_j(t),  \label{nsdeinp}
\end{align}
where the restorative dynamics of the input are described by the nonlinear function
\begin{align}
G_j (\bd) = \frac{\int_{\Rset^2} f'(U(\x)) U_{x_j} ( \x ) \tilde{I}(\x + \bd ) d \x}{\int_{\Rset^2} f'(U(\x)) U_{x_j}^2(\x ) d \x}, \ \ \ \  \ j \in \{ 1,2 \},  \label{Gintj}
\end{align}
and spatiotemporal noise provides the effective noise perturbations to the bump position through the white noise terms
\begin{align}
{\mc W}_j(t) =  \frac{\int_{\Rset^2} f'(U(\x)) U_{x_j} ( \x ) W(\x , t ) d \x}{\int_{\Rset^2} f'(U(\x)) U_{x_j}^2(\x ) d \x}, \ \ \ \  \ j \in \{ 1,2\}.
\end{align}
Note that $\langle d {\mc W}_j(t) \rangle = 0$ and $\langle d {\mc W}(t) d {\mc W}(s) \rangle = 2 D_j \delta (t-s) dt ds$ with $D_j$ given by (\ref{effdiff}). As we demonstrated in our analysis of stimulus-driven bump existence and stability, $\bar{\bd}  = (0,0)$ is a stable fixed point of the noise-free system $\dot{\Delta}_j(t) = - \ve^{1/2} G_j(\bd (t))$; $j=1,2$. Linearizing about this solution yields the multivariate OU process
\begin{align}
d \Delta_j(t) + \ve^{1/2} \beta_j \Delta_j(t) d t = \ve^{1/2} d {\mc W}_j(t), \label{inpou}
\end{align}
where
\begin{align}
\beta_j = G'(0) = \frac{\int_{\Rset^2} f'(U(\x)) U_{x_j} ( \x ) \tilde{I}_{x_j}(\x) d \x}{\int_{\Rset^2} f'(U(\x)) U_{x_j}^2(\x ) d \x}, \ \ \ \  \ j \in \{ 1,2 \},
\end{align}
assuming $\int_{\Rset^2} f'(U(\x)) U_{x_j} ( \x ) \tilde{I}_{x_k}(\x) d \x \equiv 0$, when $j \neq k$, which is the case when $I( \x )$ is even symmetric along the $x_1$ and $x_2$ directions as (\ref{radgauss}) and (\ref{transgauss}) are. In this case, we can use standard properties of an OU process to compute the mean $\langle \Delta_j(t) \rangle = \Delta_j(0)e^{- \ve^{1/2} \beta_j t}$ and variance
\begin{align}
\langle \Delta_j(t)^2 \rangle = \frac{ \ve^{1/2} D_j}{\beta_j} \left( 1 - e^{-2 \ve^{1/2} \beta_j t} \right),  \label{inpouvar}
\end{align}
so the variance $\langle \Delta_j(t)^2 \rangle$ will approach a constant $\ve^{1/2} D_j/\beta_j$ as $t \to \infty$ and the mean converges to the fixed point $\bar{\bd} = (0,0) $. Thus, we can describe the stochastic dynamics of the position $\bd (t)$ approximately using a multivariate OU process (\ref{inpou}) or with higher order corrections through the nonlinear SDE (\ref{nsdeinp}).

\begin{figure}[tb]
\begin{center} \includegraphics[width=6.4cm]{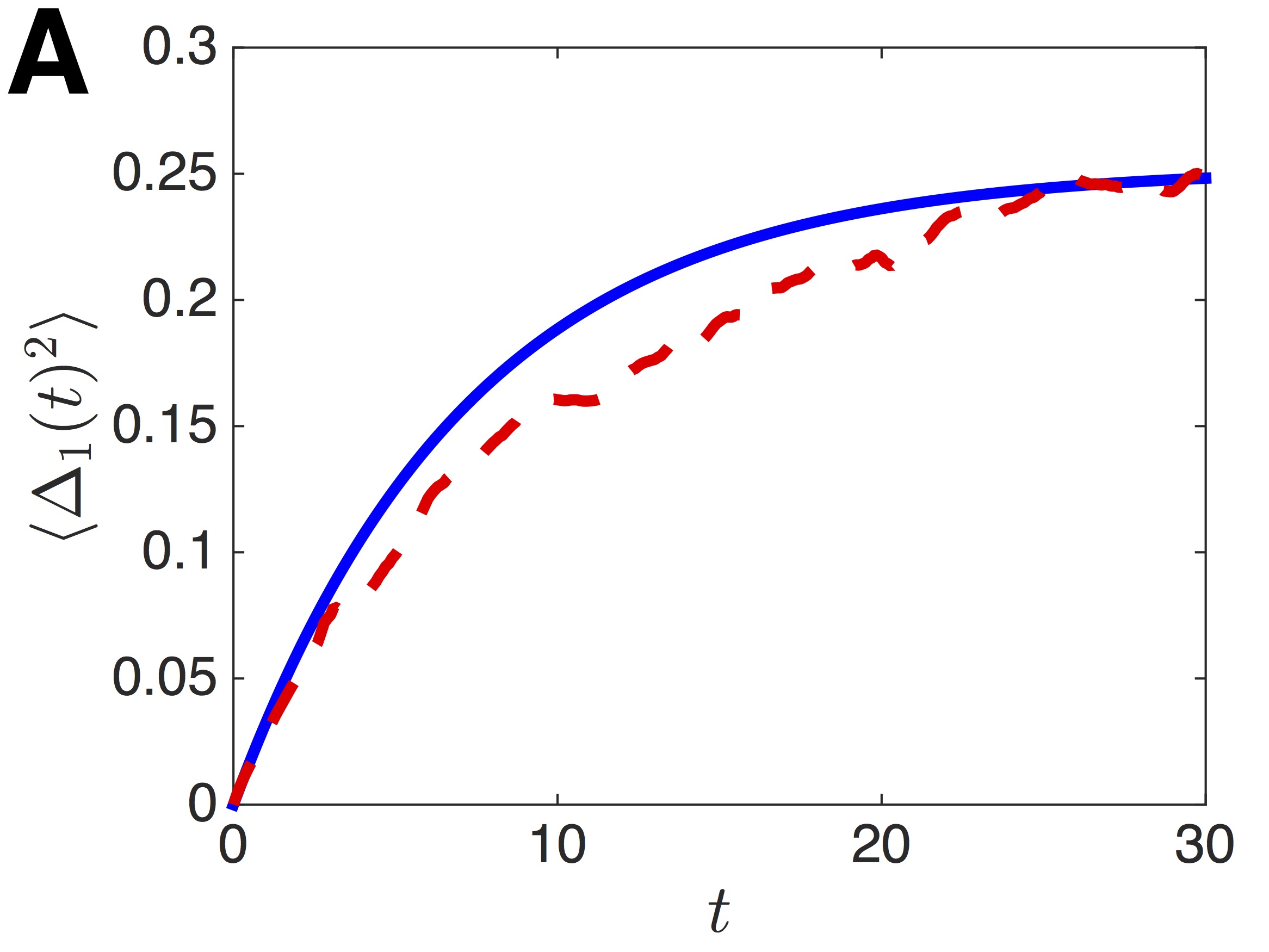} \includegraphics[width=5.6cm]{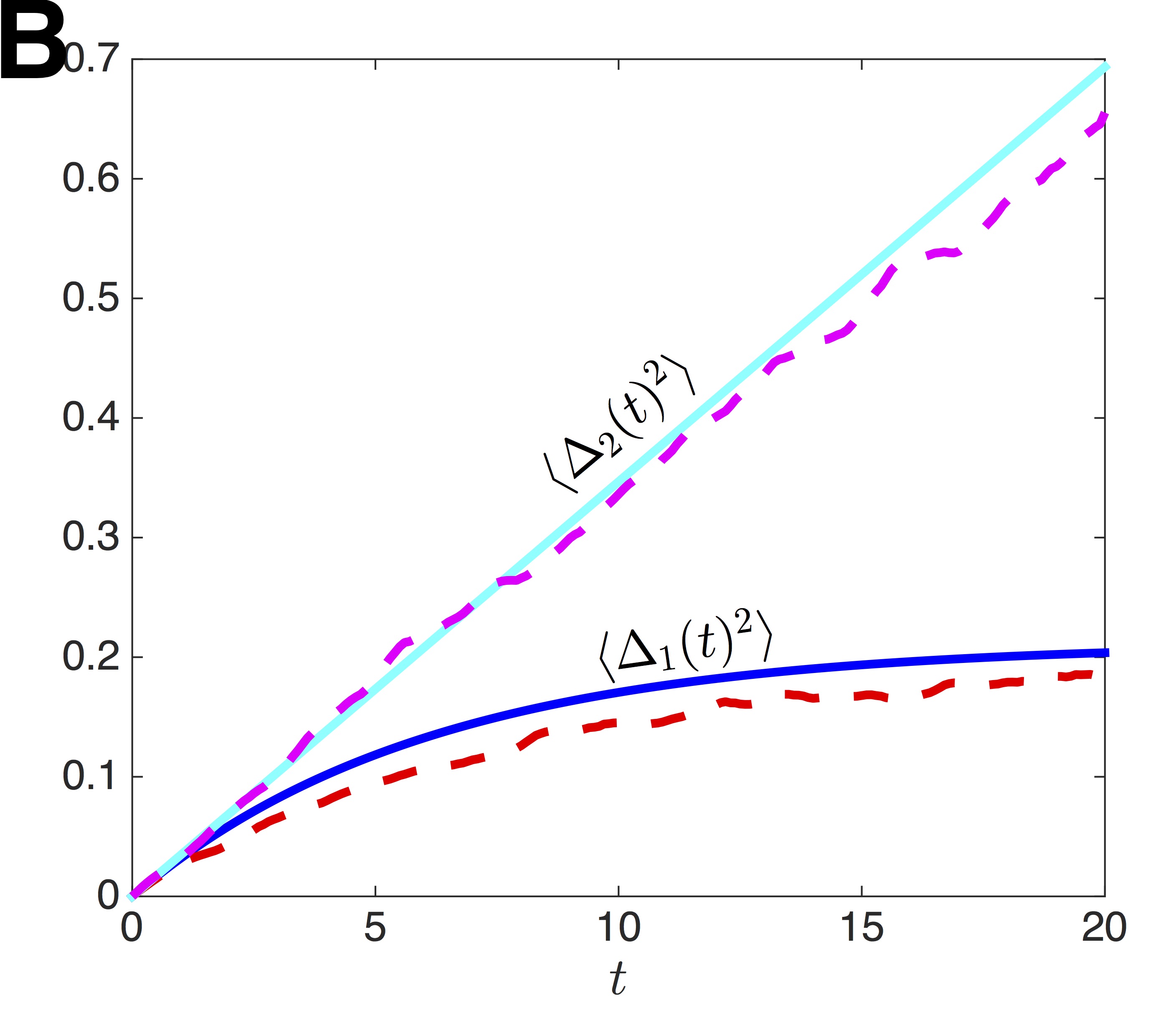}  \end{center}
\caption{Variance of the bump position $\bd (t) = ( \Delta_1(t), \Delta_2 (t))$ evolving according to the stochastic model (\ref{inplang}) with Heaviside firing rate (\ref{Hrate}) and Bessel function weight (\ref{weight}). ({\bf A}) Long-term variance $\langle \Delta_1(t)^2 \rangle$ saturates according to a multivariate Ornstein-Uhlenbeck process when input is given by a radially symmetric Gaussian (\ref{radgauss}). Results from numerical simulations (dashed line) are well matched to our theoretical result (\ref{inpouvar}). An identical picture exists for $\langle \Delta_2(t)^2 \rangle$, the variance along the $x_2$ direction. ({\bf B}) Variance along the $x_2$ direction $\langle \Delta_2(t)^2 \rangle$ climbs linearly and variance $\langle \Delta_1(t)^2 \rangle$ along the $x_1$ direction saturates when input is given by a translationally symmetric Gaussian (\ref{transgauss}).  Parameters are $\kappa = 0.2$, $A_0 = 1$, $\sigma = 2$, and $w$ is (\ref{mexweight}) with $[c_1, c_2, c_3, c_4 ]=[5/3, -5/3, -1/2, 1/2]$. Numerical calculations of variance use 1000 realizations each.}
\label{fig6}
\end{figure}

\subsection{Explicit results for the Heaviside firing rate} We can explicitly calculate the variances $\langle \Delta_j (t)^2 \rangle$ as described by the formula (\ref{inpouvar}) in the case of a Heaviside firing rate function (\ref{Hrate}), Bessel function weight kernel (\ref{weight}), and cosine noise correlations (\ref{coscorr}). Thus, the derivative $f'(U)$ is given by (\ref{fderu}), spatial derivatives $U_{x_j}$ ($j=1,2$) are given by (\ref{Udiff}), and the diffusion coefficients $D_j$ ($j=1,2$) are defined by the formula (\ref{diffjH}). Again, the impact of spatiotemporal noise in bump dynamics is primarily determined by interactions that occur at the bump boundary $r \equiv a$. Furthermore, in the case of a radially symmetric input such as the Gaussian (\ref{radgauss}), we can compute the coefficients
\begin{align}
\beta_j = \frac{\int_{\Rset^2} f'(U(\x)) U_{x_j}(\x) \tilde{I}_{x_j}(\x) d \x}{\int_{\Rset^2} f'(U(\x)) U_{x_j}^2 (\x) d \x} = \frac{a \pi \tilde{I}'(a)}{a \pi U'(a)} = \frac{\tilde{I}'(a)}{U'(a)}, \ \ \ \ \ j = \{1,2 \}.
\end{align}
Therefore, in the long time limit,  the variances $\langle \Delta_j (t)^2 \rangle$ will saturate to
\begin{align}
\lim_{t \to \infty} \langle \Delta_j(t)^2 \rangle = \frac{\ve^{1/2} D_j}{\beta_j} = \frac{4 \ve J_1(a)^2}{I'(a) U'(a)}.
\end{align}
We compare our explicit calculation of the variance to results from numerical calculations for the case of a radially symmetric Gaussian (\ref{radgauss}) in Fig. \ref{fig6}{\bf A}.

In the case of the translationally symmetric Gaussian (\ref{transgauss}), we have that $I_{x_2}(\x) \equiv 0$, so $\beta_2  \equiv 0$. As a result $\langle \Delta_2(t)^2 \rangle = \ve D_2 t$, and the variance is only mean reverting along the $x_1$ direction, as demonstrated by the numerical simulation in Fig. \ref{fig5}. The coefficient describing the systematic dynamics along the $x_1$ direction is
\begin{align}
\ve^{1/2} \beta_1 = \frac{2 a A_0 \int_0^{2 \pi} {\rm exp} \left[ \displaystyle  -a^2 \cos^2 \theta/ \sigma^2 \right] \cos^2 \theta  d \theta }{\pi \sigma^2 |U'(a)| },
\end{align}
which can be computed using quadrature to evaluate the formula (\ref{inpouvar}) for $j=1$. We compare these theoretical results with averages across numerical realizations in Fig. \ref{fig6}{\bf B}. Thus, the variance along the $x_1$-direction saturates, while the variance along the $x_2$-direction indefinitely climbs linearly.

\subsection{Statistics of the nonlinear Langevin equation} More accurate approximations of the variances $\langle \Delta_j^2 \rangle$ can be obtained by performing an analysis of the full nonlinear Langevin equation (\ref{nsdeinp}) for the stochastic motion of the input-driven bump. In \cite{bressloff15}, it was recently shown this can be particularly useful when there are multiple distinct fixed points of the noise free equations $\dot{\Delta}_j = - \ve^{1/2} G_j( \Delta )$ ($j=1,2$), as noise can eventually cause a phase-slip so a linearized approximation is no longer valid. Here, we demonstrate that a derivation of nonlinear Langevin equations can be extended to spatiotemporal patterns evolving in two-dimensions. Also, even if the position $\bd(t) = ( \Delta_1(t), \Delta_2(t))$ does remain close to a single stable fixed point, the stationary probability density $P_0( \bd )$ of (\ref{nsdeinp}) can be considerably different than that of the truncated OU process (\ref{inpou}). Thus, we briefly present the computation of this stationary probability density $P_0(\bd)$ by utilizing the associated Fokker-Planck equation, saving a more extensive study for future work.

\begin{figure}[tb]
\begin{center} \includegraphics[width=6.2cm]{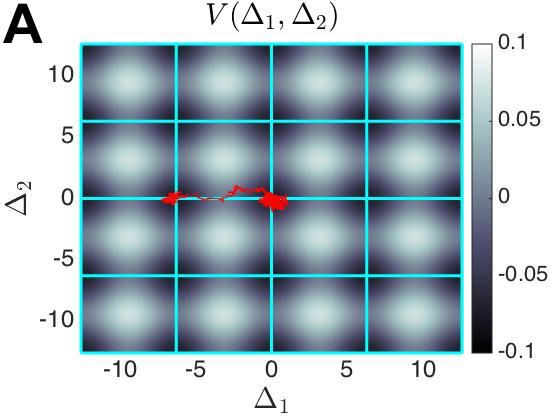}   \includegraphics[width=6.2cm]{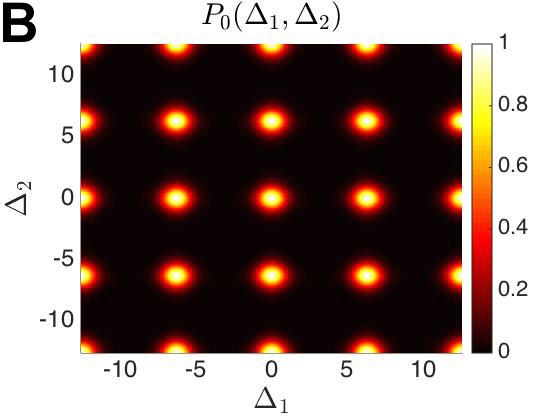}  \end{center}
\caption{Nonlinear dynamics in the input driven neural field (\ref{inplang}) on the plane $\Rset^2$ with Heaviside firing rate (\ref{Hrate}) and Bessel function weight (\ref{weight}), subject to the periodic input $\tilde{I}(\x) = 0.3(\cos x_1 + \cos x_2)$. ({\bf A}) Bump position (thin line)  sampled from a single realization of the system (\ref{inplang}) for $t\in[0,4000]$, superimposed on a plot of the potential $V(\bd)$. Over long periods of time, the stochastically-driven position of bumps tends to dwell primarily in the vicinity of minima of the potential $V(\bd)$ defined by (\ref{2dpotwell}). However, there are rare events whereby the bump transitions to a neighboring potential well. Thick lines represent the paths of least action between attractors at $(2m \pi, 2n \pi )$, $m,n \in {\mathbb Z}$. ({\bf B}) Re-normalized stationary probability density $P_0(\x)$ (so that $||P_0(\x)||_{max}=1$, otherwise $P_0(\x)$ would be infinitesimally small everywhere) has peaks at the minima of the potential function $V(\bd)$.  Parameters are $\ve = 0.025$, $\kappa = 0$, and weight $w$ is (\ref{mexweight}) with $[c_1,c_2,c_3,c_4] = [5/3, -5/3, -1/2, 1/2]$.}
\label{fig7}
\end{figure}

The stochastic dynamics is shaped by an underlying potential function $V( \bd )$, which is the solution to the pair of equations
\begin{align}
\frac{d V}{d \Delta_1} = \ve^{1/2} G_1( \bd), \hspace{5mm} \frac{d V}{d \Delta_2} = \ve^{1/2} G_2 ( \bd ),   \label{poteqns}
\end{align}
so that the attractors of the noise free system are the minima of $V( \bd )$. While (\ref{poteqns}) cannot always be solved explicitly, we will present an example below where they can, for illustration. To analyze the system, we reformulate (\ref{nsdeinp}) as an equivalent Fokker-Planck equation \cite{gardiner04}
\begin{align}
\frac{\pd P( \bd ,t)}{\pd t} = \sum_{j=1}^2 \left\{ \frac{\pd}{\pd \Delta_j} \left[  \ve^{1/2} G_j (\bd ) P( \bd , t) \right] + \ve D_j \frac{\pd^2 P( \bd , t)}{\pd \Delta_j^2} \right\},
\end{align}
where $P(\bd,t)$ is the probability of finding the bump at position $\bd = (\Delta_1, \Delta_2)$ at time $t$. Assuming there is a single stable attractor defined by the potential $V( \bd )$, in the long time limit
\begin{align}
\lim_{t \to \infty} P( \bd ,t ) = P_0( \bd ) = \chi \exp \left[ - \frac{V( \bd )}{\ve D} \right],  \label{nlstatdens}
\end{align}
in the case of rotationally symmetric noise $D_1=D_2=D$, where $\chi$ is a normalization factor such that $\int_{\Rset^2} P_0( \bd ) d \bd = 1$. The long time variance $\langle || \bd||^2 \rangle = \langle \Delta_1^2 \rangle + \langle \Delta_2^2 \rangle$ is thus given by the integral
\begin{align}
\langle \Delta_1^2 \rangle + \langle \Delta_2^2 \rangle = \int_{\Rset^2} (\Delta_1^2 + \Delta_2^2) P_0( \bd ) d \bd.
\end{align}

To demonstrate our analysis on a specific example, we focus on an input which allows explicit computation of the steady state distribution. Thus, we take the external input $\tilde{I} ( \x ) = A_0(\cos x_1 + \cos x_2)$ and assume we then wish to compute the statistics of the stationary probability density $P_0( \bd )$. Furthermore, we take the Heaviside firing rate function (\ref{Hrate}), then the integrals (\ref{Gintj}) can be simplified by making the substitutions $x_1 = r \cos \theta$ and $x_2 = r \sin \theta$ and integrating out the radial coordinate $r$, so
\begin{align}
G_1 ( \bd ) &= \frac{U'(a) \int_0^{2 \pi} \cos ( \theta ) \tilde{I}( \ab + \bd ) d \theta}{U'(a)^2 \int_0^{2 \pi} \cos^2 \theta d \theta}  = \frac{1}{U'(a) \pi} \int_0^{2 \pi} \cos (\theta) \tilde{I}( \ab + \bd ) d \theta \\
G_2 ( \bd ) &= \frac{U'(a) \int_0^{2 \pi} \sin ( \theta ) \tilde{I} ( \ab + \bd ) d \theta}{U'(a)^2 \int_0^{2 \pi} \sin^2 \theta d \theta} = \frac{1}{U'(a) \pi} \int_0^{2 \pi} \sin \theta \tilde{I}( \ab + \bd ) d \theta,
\end{align}
where $\ab = (a, \theta)$. Selecting the doubly periodic function for our external input $\tilde{I} ( \x ) = A_0(\cos x_1 + \cos x_2)$, we generate terms similar to those that arose in our explicit calculation of the diffusion coefficient $D_j$ (\ref{diffjH}). Subsequently, we find we can evaluate these explicitly, to arrive at the compact expression
\begin{align}
G_j ( \bd ) = - \frac{2 A_0 J_1(a)}{U'(a)} \sin ( \Delta_j), \ \ \ \ \ \{j=1,2 \}.
\end{align}
Thus, the positions $\Delta_1$ and $\Delta_2$ evolve independently in nonlinear system (\ref{nsdeinp}), in the case of this specific input function. Thus, it is straightforward to evaluate the potential function as the solution to (\ref{poteqns}), finding
\begin{align}
V( \bd ) = - \frac{2 \ve^{1/2} A_0 J_1(a)}{|U'(a)|} \left[ \cos (\Delta_1) + \cos( \Delta_2 ) \right], \label{2dpotwell}
\end{align}
which can then be utilized to compute the statistics of the stationary probability density (\ref{nlstatdens}). We demonstrate that stochastic trajectories of the bump tend to dwell mostly in the minima of the potential function defined by (\ref{2dpotwell}) in Fig. \ref{fig7}{\bf A}. Such durations are interrupted by abrupt transitions of the bump between neighboring wells as in the one-dimensional case \cite{kilpatrick13}. An example of the rescaled stationary probability density is given in Fig. \ref{fig7}{\bf B}. We save a more thorough analysis of these results for subsequent work.

\section{Discussion} \label{disc}

We have analyzed the impact of additive noise on the stochastic motion of bumps in planar neural field equations. In networks with no spatial heterogeneity, noise causes bumps to wander according to two-dimensional Brownian motion. The diffusion coefficient associated with this motion can be approximated using an asymptotic expansion that treats the impact of noise perturbatively. Assuming the bump retains its profile, to first order, we can derive an effective diffusion equation for the bump's position as a function of time. Notably, the dynamics of the bumps can be separated into diffusion along the canonical directions ($x_1,x_2$) in $\Rset^2$. In the presence of spatially heterogeneous external inputs, the stochastic bumps no longer obey dynamics well described by pure diffusion. Rather, bumps are attracted to the local maxima of the input functions, so their motion can be approximated by multivariate Ornstein-Uhlenbeck processes. In particular, we find that the geometry of the external inputs define the manifold to which bumps are attracted. Radially symmetric inputs attract bumps to a single point and translationally symmetric peaked inputs attract bumps to a one-dimensional line through $\Rset^2$.

Our primary motivation for exploring these models comes from extensive experimental literature on persistent activity representing spatial working memory \cite{funahashi89,mcnaughton06}. Particularly pertinent to this work are place cell and grid cell networks, which track a mammal's idiothetic position in two-dimensional space as it navigates through its environment \cite{moser08}. Networks in cortex and hippocampus are capable of encoding analog spatial variables for time periods lasting seconds up to tens of minutes on length scales of hundreds of meters \cite{hafting05}. The specific single neuron and network architectural features that engender this impressive accuracy are the subject of ongoing research \cite{burak09}. We will extend the results presented here in future work by addressing how spatially heterogeneous architecture\cite{kilpatrick13,kilpatrick13c} and the known multilayered structure of grid cell networks \cite{hafting05,brun08} may contribute to the observed precision of spatial navigation networks of the brain. Thus, we will extend the methods we have developed for analyzing coupling between multiple layers of one-dimensional stochastic neural fields to two-dimensions \cite{kilpatrick13b,bressloff15}. Furthermore, it is important to note that synaptic connections projecting from long-range axons can be subject to axonal transmission delays, as in the planar neural field study of Hutt and Rougier \cite{hutt10}. In future work, we plan to explore the impact of delays within and between layers of planar neural fields on the stochastic motion of bumps, as we did in our recent work on one-dimensional systems \cite{kilpatrick15}.  

Lastly, we aim to extend this work by further analyzing the impact of external inputs on the stochastic dynamics of bumps in planar neural fields. As we found here, the local maxima of external inputs determine the most likely spatial regions to find bumps that are perturbed by spatiotemporal noise. However, this general picture may change if we consider external inputs that vary in time. For instance, stationary external inputs that turn on and off periodically will likely not pin bumps in place as well. In a similar way, we could explore how well bumps track the positions inputs that move smoothly in space, dependent on their speed. It is possible that noise may actually improve bumps' tracking of external input positions, in some parameter regimes. We may be able to exploit relative slow or fast timescales of external inputs to extend our presented asymptotic analysis, which assumes inputs are stationary.

\bibliography{2dbump}
\bibliographystyle{siam}
\end{document}